\documentclass[11pt]{article}
\usepackage{tikz}
\usepackage{colortbl}
\usetikzlibrary{tikzmark}
\usepackage{caption}
\usepackage{mdframed}

\usepackage{epsfig,subfigure}
\usepackage{amssymb}
\usepackage[fleqn]{amsmath}
\usepackage{color}
\usepackage{blkarray}

\usepackage{arydshln}
\newcommand{\PIRA}{{\color{black} \it PIR1}}

\newcommand{\PIRB}{{ \color{black} \it PIR2}}

\newcommand{\PC}{{\color{black} \it PC}}

\def\QED{\mbox{\rule[0pt]{1.5ex}{1.5ex}}}

\def\new{\mbox{\texttt{new}}}

\usepackage{graphicx}
\usepackage{color}

\usepackage{cite}
\definecolor{armygreen}{rgb}{0.29, 0.33, 0.13}

\usepackage{amssymb}
\usepackage{amsmath,amsfonts,amssymb}
\usepackage{verbatim}
\usepackage{stfloats}
\usepackage[bookmarks=false]{}
\usepackage{algorithm}
\usepackage{algcompatible}

\newtheorem{theorem}{Theorem}

\newtheorem{lemma}{Lemma}

\newcommand\m[1]{%
\mbox{\small #1}%
}

\newcommand\blfootnote[1]{%
  \begingroup
  \renewcommand\thefootnote{}\footnote{#1}%
  \addtocounter{footnote}{-1}%
  \endgroup
}

\begin{document}
\date{}
\title{
The Capacity of Private Computation
}
\author{ \normalsize Hua Sun and Syed A. Jafar 
}
\maketitle
\blfootnote{Hua Sun (email: hua.sun@unt.edu) is with the Department of Electrical Engineering at the University of North Texas. Syed A. Jafar (email: syed@uci.edu) is with the Center of Pervasive Communications and Computing (CPCC) in the Department of Electrical Engineering and Computer Science (EECS) at the University of California Irvine. 
}
\begin{abstract}
We introduce the problem of private computation, comprised of $N$ distributed and non-colluding servers, $K$ independent datasets, and a user who wants to compute a function of the datasets privately, i.e., without revealing which function he wants to compute, to any individual server. This private computation problem is a strict generalization of the private information retrieval (PIR) problem, obtained by expanding the PIR message set (which consists of only independent messages) to also include functions of those messages. The capacity of private computation, $C$, is defined as the maximum number of bits of the desired function that can be retrieved per bit of total download from all servers. 
We characterize the capacity of private computation, for $N$ servers and $K$ independent datasets that are replicated at each server, when the functions to be computed are arbitrary linear combinations of the datasets. Surprisingly, the capacity, $C=\left(1+1/N+\cdots+1/N^{K-1}\right)^{-1}$, matches the capacity of PIR with $N$ servers and $K$ messages. 
Thus, allowing arbitrary linear computations does not reduce the communication rate compared to pure dataset retrieval. 
The same insight is shown to hold even for arbitrary non-linear computations when the number of datasets $K\rightarrow\infty$.
\end{abstract}

\newpage
\allowdisplaybreaks
\section{Introduction}

Motivated by privacy concerns in distributed computing applications, we introduce the private computation (PC) problem, where a user wishes to privately compute a function of datasets that are stored at distributed servers. Specifically, $K$  datasets  are stored at $N$  non-colluding servers, and a user  wishes to compute a function of these datasets. A private computation scheme allows the user to compute his desired function, while revealing no information to any individual server about the identity of the desired function. The achievable rate of a private computation scheme is the ratio of the number of bits of the desired function that the user is able to retrieve, to the total number of bits downloaded from all servers. The capacity of private computation is  the supremum of achievable rates. 

The private computation problem is a strict generalization of the private information retrieval (PIR) problem, where one of the $K$ datasets is desired by the user, i.e., the function to be computed simply returns the desired dataset. The capacity was characterized recently for PIR in \cite{Sun_Jafar_PIR} and for several of its variants in \cite{Sun_Jafar_TPIR, Sun_Jafar_SPIR, Banawan_Ulukus, Wang_Skoglund, Banawan_Ulukus_Byzantine, Tandon_CachePIR}. In the PIR setting, the datasets are called messages and all messages are independent. Private computation may also be viewed as  PIR with \emph{dependent} messages, where each possible function that may be desired by a user is interpreted as a dependent message, i.e., a message whose value depends on other messages. 

Our main result is the  characterization of  the capacity of  
private computation, where a user wishes to compute arbitrary linear combinations of $K$ independent datasets (messages), replicated at $N$ servers. Note that if the user can only choose one of  $M=K$ independent linear combinations, then the setting is equivalent to the PIR problem with $K$ messages and $N$ servers. From \cite{Sun_Jafar_PIR}, we know that the capacity of PIR in this setting is equal to $\left(1+1/N+\cdots+1/N^{K-1}\right)^{-1}$. Surprisingly, we show that even if the user wishes to compute arbitrary linear combinations of the $K$ datasets, the capacity of private computation remains $\left(1+1/N+\cdots+1/N^{K-1}\right)^{-1}$, i.e., in terms of capacity, arbitrary linear computation incurs no additional penalty. 

The capacity achieving scheme for private computation that is presented in this work is a highly structured adaptation of the   capacity achieving scheme for PIR that was introduced in \cite{Sun_Jafar_PIR}.  Specifically, the private computation scheme utilizes an optimized symbol index structure, and a sophisticated assignment of signs (`$+$' or `$-$') to each symbol in order to optimally exploit the linear dependencies. {\color{black}A surprising feature of the optimal private computation scheme is that the query construction does not depend on the linear combining coefficients that define the set of possible functions that may be computed by the user.  }

{\color{black}Finally, we note that following the ArXiv posting of our  capacity results for the elemental setting of private computation with $N=2, K=2$, arbitrary $M$ (first version of this paper, posted October 30, 2017), an independent work on `\emph{private function retrieval}' was posted on ArXiv by Mirmohseni and Maddah-Ali (reference \cite{Mirmohseni_Maddah}, posted November 13, 2017). Since the private function retrieval problem is identical to the private computation problem, it is worthwhile to compare and contrast the two works. To this end, we note that while  there is no overlap in the achievable schemes proposed in the two works, the general capacity result presented in this paper  subsumes and strictly improves upon the results of \cite{Mirmohseni_Maddah}.  In particular,  \cite{Mirmohseni_Maddah} presents two  results. The first result of \cite{Mirmohseni_Maddah} is a capacity characterization of private computation when $N=2$,   $K$ is arbitrary, and  the set of functions that may be computed is comprised of all possible linear combinations of the $K$ message sets --- albeit limited to binary coefficients. This result is recovered as a special case of our general capacity result in this paper. In this case, although the achievable schemes of \cite{Mirmohseni_Maddah} and this work are different, they both achieve capacity. The second result of \cite{Mirmohseni_Maddah} is an extension of their achievable scheme to  general $N, K$  and non-binary combining coefficients, although the optimality of the achievable scheme is left open. For this general case, our capacity characterization implies that the achievable scheme of \cite{Mirmohseni_Maddah} is strictly suboptimal.}

{\it Notation: 
For  integers $Z_1, Z_2, Z_1 \leq Z_2$, we use the compact notation $[Z_1:Z_2]=\{Z_1, Z_1+ 1,\cdots, Z_2\}$. 
For an index set $\mathcal{I} = \{i_1, i_2 \cdots, i_k\}$, the notation $A_{\mathcal{I}}$ represents the set $\{A_i, i \in \mathcal{I}\}$.
The notation $X \sim Y$ is used to indicate that $X$ and $Y$ are identically distributed. 
For a matrix ${\bf A}$, ${\bf A}^{T}$ represents its transpose and $|{\bf A}|$ represents its determinant. For a set $S$, $|S|$ represents its cardinality. 
For sets $S_1, S_2$, we define $S_1/S_2$ as the set  of elements that are in $S_1$ and not in $S_2$. 
}

\section{Problem Statement and Definitions}\label{sec:model}
Consider the private computation problem with $N$ servers and $K$ datasets. We will assume that the datasets are replicated at all servers, that the servers do not collude, and that the functions to be computed are linear combinations of the messages. We will focus primarily on this basic setting which opens the door to  numerous other open problems through various generalizations, e.g.,  
coded storage instead of replication, 
colluding servers, symmetric privacy requirements, non-linear functions, etc.

The $K$ datasets, denoted by $W_{d_1}, \cdots, W_{d_K} \in\mathbb{F}_p^{L\times 1}$, are each comprised of $L$ i.i.d. uniform symbols from a finite field $\mathbb{F}_p$.  In $p$-ary units,
\begin{eqnarray}
H(W_{d_1}) &=& \cdots = H(W_{d_K}) = L, \\
H(W_{d_1}, \cdots, W_{d_K}) &=& H(W_{d_1}) + \cdots + H(W_{d_K}).
\end{eqnarray}

A linear combination of these datasets is represented as a dependent message,
\begin{eqnarray}
W_m = {\bf v}_{m} [W_{d_1}, \cdots, W_{d_K}]^T = v_{m(1)} W_{d_1} + \cdots + v_{m(K)} W_{d_K}, m \in [1:M] \label{linear}
\end{eqnarray}
where ${\bf v}_{m} = [v_{m(1)}, \cdots, v_{m(K)}]$ consists of $K$ constants from $\mathbb{F}_p$, and `$+$' represents element-wise addition over $\mathbb{F}_p$. 
Without loss of generality, we assume $M \geq K$ and 
$[{\bf v}_1; {\bf v}_2; \cdots; {\bf v}_K] = {\bf I}_K$, where ${\bf I}_K$ is the $K \times K$ identity matrix. Thus, $(W_1, W_2, \cdots, W_K)=(W_{d_1}, W_{d_2}, \cdots, W_{d_K})$.

There are $N$ servers and each server stores all datasets $W_{d_1}, \cdots, W_{d_K}$. 
A user privately generates $\theta \in [1:M]$ and wishes to compute (retrieve) $W_\theta$ while keeping $\theta$ a secret from each server. Depending on $\theta$, there are $M$ strategies that the user could employ to privately compute his desired function. For example, if $\theta = m$, then in order to compute $W_m$, the user employs $N$ queries, $Q_1^{[m]}, \cdots, Q_N^{[m]}$. Since the queries are determined by the user with no knowledge of the realizations
of the messages, the queries must be independent of the messages,\footnote{The message sets $(W_{d_1}, \cdots, W_{d_K})$ and $(W_1, W_2, \cdots, W_M)$ are invertible functions of each other, so, e.g., conditioning on one is the same as conditioning on the other.
 }
\begin{eqnarray}
\forall m \in [1:M], I(W_1, \cdots, W_M; Q_1^{[m]}, \cdots, Q_N^{[m]}) = 0. \label{qmind}
\end{eqnarray}
The user sends $Q_n^{[m]}, n \in [1:N]$ to the $n^{th}$ server. Upon receiving $Q_n^{[m]}$, the $n^{th}$ server generates an answering string $A_n^{[m]}$, which is a function of $Q_n^{[m]}$ and the data stored (i.e., all the messages),
\begin{eqnarray*}
\forall m \in [1:M], n\in[1:N], H(A_n^{[m]} | Q_n^{[m]}, W_1, \cdots, W_M) = 0. \label{ans_det}
\end{eqnarray*}
Each server returns to the user its answer $A_n^{[m]}$. From all the information that is now available to the user $(A_1^{[m]}, \cdots, A_N^{[m]}, {\color{black} Q_1^{[m]}, \cdots, Q_N^{[m]}})$,  the user decodes the desired message $W_m$ according to a decoding rule that is specified by the private computation scheme. Let $P_e$ denote the probability of error achieved with the specified decoding rule.

To protect the user's privacy, the $M$ strategies must be indistinguishable (identically distributed) from the perspective of each server, i.e., the following privacy constraint must be satisfied $\forall n\in[1:N],\forall m\in[1:M]$,
\begin{eqnarray}
\mbox{[Privacy]} 
~~ (Q_n^{[1]}, A_n^{[1]}, W_1, \cdots, W_M) \sim (Q_n^{[m]}, A_n^{[m]}, W_1, \cdots, W_M). 
\label{privacy}
\end{eqnarray}

The PC \emph{rate} characterizes how many symbols of desired information are computed per downloaded symbol, and is defined as follows. 
\begin{eqnarray}
R \triangleq \frac{L}{D} \label{eta_def} 
\end{eqnarray}
where $D$ is the expected value (over random queries) of the total number of symbols downloaded by the user from all servers. 

A rate $R$ is said to be $\epsilon$-error achievable if there exists a sequence of private computation schemes, indexed by $L$, each of rate greater than or equal to $R$, for which $P_e\rightarrow 0$ as $L\rightarrow\infty.$ Note that for such a sequence of private computation schemes, from Fano's inequality, we have
\begin{eqnarray}
 \mbox{[Correctness]} ~~ H(W_m | A_1^{[m]}, \cdots, A_N^{[m]}, {\color{black} Q_1^{[m]}, \cdots, Q_N^{[m]}}) = o(L) \label{corr}
\end{eqnarray}
where any function of $L$, say $f(L)$, is said to be $o(L)$ if $\lim_{L\rightarrow\infty} f(L)/L = 0$.
The supremum of $\epsilon$-error achievable rates is called the capacity $C$. 

\section{Capacity of Private Computation}
Theorem \ref{thm} states our main result.
\begin{theorem}\label{thm}
For the private computation problem where a user wishes to privately retrieve one of $M$ arbitrary\footnote{Note that $M\geq K$ and the $M$ {\color{black}linear combinations} contain $K$ linearly independent ones, so that $H(W_1,W_2,\cdots, W_M)=H(W_{d_1},W_{d_2}, \cdots, W_{d_K})=KL$.} linear combinations of $K$ independent datasets from $N$ servers, the capacity is $C=\left(1+1/N+\cdots+1/N^{K-1}\right)^{-1}$. 
\end{theorem}

When $M = K$, the problem reduces to the PIR problem with $N$ servers and $K$ messages, for which the capacity is $\left(1+1/N+\cdots+1/N^{K-1}\right)^{-1}$ \cite{Sun_Jafar_PIR}. Adding more computation requirements $M > K$ can not help (surprisingly it does not hurt either), so the converse of Theorem \ref{thm} is implied. We only need to prove the achievability, which is presented in Section \ref{sec:a2}. 

It is quite surprising that increasing the number of messages by including arbitrary linear combinations of $K$ datasets does not reduce capacity for all linear computation settings. A natural question then is whether this insight holds more broadly. Remarkably, the insight is also true for arbitrary non-linear computations, when the number of datasets is large ($K \rightarrow \infty$). It turns out that in this case, again the capacity of private computation is equal to the capacity of PIR. This supplemental result is  rather straightforward and is stated in the following theorem.
\begin{theorem}\label{thm2}
For the private computation problem with $K$ independent datasets, $W_{k}$, $k\in[1:K]$, $H(W_k)=L$,    arbitrary $N$ servers and $M-K$ arbitrary (possibly non-linear) dependent messages,  $W_m$, $m\in[K+1:M]$, $H(W_m \mid W_{k}, k\in[1:K])=0$, $H(W_m)\leq L$, if $K \rightarrow \infty$, then the capacity of private computation  $C \rightarrow 1 - 1/N$, which is the capacity of PIR with $K\rightarrow\infty$ messages and $N$ servers.
\end{theorem}
{\it Proof:} For Theorem \ref{thm2}, the achievability is identical to the symmetric PIR\footnote{ Theorem \ref{thm2} extends immediately to the symmetric private computation problem, where the user is prohibited from learning anything beyond the desired function.}
 scheme of Theorem 1 in \cite{Sun_Jafar_SPIR} (see also \cite{PIRfirstjournal, Shah_Rashmi_Kannan}), where the $M$ functions are viewed as the messages in the symmetric PIR problem and common randomness is not used. The  dependence of the messages has no impact on privacy or correctness of that scheme. The converse follows from the converse of regular PIR \cite{Sun_Jafar_PIR} because restricting the message set to $W_k, k\in[1:K]$ cannot reduce capacity.  
\hfill\QED

\section{The Achievable Scheme }\label{sec:a2}

The private computation scheme needed for Theorem \ref{thm} builds upon and significantly generalizes the capacity achieving PIR scheme presented in \cite{Sun_Jafar_PIR, Sun_Jafar_PIRL}. If we ignore the dependence of the messages in the private computation problem and directly use the PIR scheme (capacity achieving for independent messages) in \cite{Sun_Jafar_PIR}, the rate achieved is $\left(1 + 1/N + \cdots + 1/N^{M-1} \right)^{-1}$, which is strictly less than $\left(1 + 1/N + \cdots + 1/N^{K-1} \right)^{-1}$ (independent of $M$), the capacity of private computation. To optimally exploit the dependence of the messages, we start with the original PIR scheme of \cite{Sun_Jafar_PIR} and incorporate two new ideas. For ease of reference, let us denote the original PIR scheme of \cite{Sun_Jafar_PIR} as $\PIRA$. 
\begin{enumerate}
\item[(1)] {\bf Index assignment:} Additional structure is required from symbol indices within the queries because dependence only exists across message symbols associated with the same index. This requirement yields  a new PIR scheme, that we will denote as $\PIRB$. If the messages are independent, then  in terms of downloads $\PIRB$  is as efficient as $\PIRA$, i.e., they are both capacity achieving schemes. 

\item[(2)] {\bf Sign assignment:} The index structure of $\PIRB$ seems essential to accommodate dependent messages. By itself, however, it is not sufficient.\footnote{\color{black}Remarkably, if the field $\mathbb{F}_p$ in (\ref{linear}) is restricted to $\mathbb{F}_2$ then  $\PIRB$ is sufficient to achieve the capacity of private computation. This is because sign-assignments are redundant over $\mathbb{F}_2$, i.e., $+x$ and $-x$ are equivalent over $\mathbb{F}_2$.} For example, the queries in both $\PIRA$ and $\PIRB$ are comprised of \emph{sums} of symbols. Depending on the form of message dependencies, more sophisticated forms of combining symbols within queries may be needed. For our present purpose, with linear message dependencies, we will need both sums and differences. To this end, we need to carefully assign a `sign' (`$+$' or `$-$') to each symbol. The sign assignment produces the optimal private computation scheme, denoted $\PC$, for Theorem \ref{thm}. 
\end{enumerate}

To present these schemes, we need to introduce the following notation.
Let $\pi$ represent a permutation over $[1:L]$. For all $m\in[1:M], i\in[1:L]$ let 
\begin{eqnarray}
u_m(i)&=&\sigma_i W_m(\pi(i))
\end{eqnarray}
Thus, $W_m(\pi(i))$ are the symbols from message $W_m$, permuted by $\pi$, and $u_m(i)$ are the corresponding signed versions obtained by scaling with $\sigma_i\in\{+1,-1\}$. Since both $m$ and $i$ are indices in $u_m(i)$, if there is a potential for confusion, we will refer to $m$ as the `message index' and $i$ as the `symbol index'. 
Note that the same permutation is applied to all messages, and the same sign variable $\sigma_i$ is applied to symbols from different messages that have the same symbol index.  Both $\pi$ and $\sigma_i$ are generated privately, independently and uniformly by the user such that they are not known to the servers. 

We will refer to the message $W_m$ equivalently as the message $u_m$. To illustrate the key ideas we will use the special $K=2, M=4, N=2$ setting as our running example in this work. 

\bigskip

\begin{mdframed}
\subsubsection*{Example $A$}
Suppose the $M = 4$ functions on the $K=2$ datasets that we wish to compute over $N = 2$ servers are the following.
\begin{eqnarray}
&& W_1 = W_{d_1} \notag\\
&& W_2 = W_{d_2}  \notag\\
&& W_3 = {\color{black}v_3}W_{d_1} + {\color{black}v'_3}W_{d_2}\notag\\
&&W_4 = {\color{black}v_4} W_{d_1} + {\color{black} v'_4}  W_{d_2} \label{eq:r1}
\end{eqnarray}
Each message consists of $L = N^M = 16$ symbols from $\mathbb{F}_p$. The specialized setting allows us to use a simpler notation as follows. 
\begin{eqnarray*}
(a_i, b_i, c_i, d_i) = (u_1(i), u_2(i), u_3(i), u_4(i))
\end{eqnarray*}
The notation is simpler because we only have symbol indices. Message indices are not necessary in this toy setting because a different letter is used for each message.
\end{mdframed}
 \bigskip

We will start with the query structure of the PIR scheme, which we will modify using the two principles outlined earlier, to obtain the private computation scheme. First we explain the index assignment step.

\subsection{Index Assignment: $\PIRB$}
In this section, we introduce the $\PIRB$ scheme, built upon $\PIRA$ by an index assignment process. The index assignments are necessary because unlike $\PIRA$ where independent permutations are applied to symbols from each message, in $\PIRB$ the same permutation is applied to symbols from every message.  For ease of exposition, we will first illustrate the index assignment process through Example $A$, and then present the general algorithm for arbitrary $K, M, N$. Since we do not use sign assignments in  $\PIRB$, the $\sigma_i$ are redundant for this scheme. Without loss of generality, the reader may assume $\sigma_i=1$ for all $i$ for $\PIRB$.
\subsubsection{{\bf Example $A$}}
Suppose the desired message is $W_1$, i.e., $\theta = 1$.
Recall the query structure of $\PIRA$, where we have left some of the indices of undesired symbols undetermined.
\begin{eqnarray*}
\begin{array}{c}
\theta=1\\
\begin{array}{|c|c|c|c|c|}\hline
\mbox{\small Server 1}&\mbox{\small Server 2}\\\hline
 a_1,  b_1,  c_1,  d_1& a_2,  b_2,  c_2,  d_2\\
 a_3 +  b_2 &  a_6 + b_1 \\
 a_4 + c_2 &  a_7 + c_1 \\
 a_5 + d_2 &  a_8 + d_1 \\
 b_* + c_* &  b_* + c_*\\
 b_* + d_* &  b_* + d_* \\
 c_* + d_* &  c_* +  d_* \\
 a_9 +  b_* +  c_* &   a_{12} + b_* +  c_*\\
 a_{10} + b_* +  d_* &   a_{13} +  b_* +  d_*\\
 a_{11} +  c_* +  d_* &  a_{14} + c_* +  d_* \\
 b_* + c_* +  d_* &   b_* + c_* +  d_* \\
 a_{15} + b_*+ c_* +  d_* &  a_{16} +  b_* +  c_* +  d_*\\
\hline
\end{array}
\end{array}
\end{eqnarray*}
Note that the first row of the query to  {\color{black}S}erver $n$, $n\in\{1,2\}$, is $a_n, b_n, c_n, d_n$, just as in $\PIRA$. In $\PIRA$, the permutations are chosen independently for each message, so that $c_n, d_n$ are not necessarily functions of $a_n, b_n$. However, here, because  we apply the same permutation to every message, and because the same sign $\sigma_n$ is applied to $a_n, b_n, c_n, d_n$, the dependence of messages is preserved in these symbols. In particular, $c_n=v_3a_n+v'_3b_n$, $d_n=v_4a_n+v'_4b_n$, and $H(a_n, b_n, c_n, d_n) = 2$ $p$-ary units.  

The next three rows of the queries to each server are $2$-sums (i.e., sums of two symbols) that are also identical to $\PIRA$, because these queries exploit the side-information from the other server to retrieve new desired symbols. However, notice that because permutations of message symbols are identical, there is a special property that holds here that is evident to each server. For example, Server $1$ notes that the $2$-sums that contain $a_i$ symbols, i.e., $a_3+b_2, a_4+c_2, a_5+d_2$ have the same index for the other symbol, in this case the index $2$. Since we do not wish to expose the identity of the desired message, the same property must hold for all messages. This observation forces the index assignments of all remaining $2$-sums. 

For example, let us consider the next query term, $b_*+c_*$, from, say, Server $1$. Since $b_2$ was mixed with $a_3$ in the query $a_3+b_2$, all $2$-sums that include some $b_i$ must have index $3$ for the other symbol. Similarly, since $c_2$ was mixed with $a_4$, all $2$-sums that include some $c_j$ must have index $4$ for the other symbol. Thus, for Server $1$, the only index assignment possible for query $b_*+c_*$ is $b_4 + c_3$. Similarly, the $b_*+d_*$ must be $b_5+d_3$ and $c_*+d_*$ must be $c_5+d_4$. All indices for $2$-sums are similarly assigned for Server $2$ as well. Thus all indices for $2$-sums are settled. 

Now let us consider $3$-sums. The index assignments for the first three rows for the $3$-sums are again straightforward, because as in \cite{Sun_Jafar_PIR}, these are side-information exploitation terms, i.e., new desired message symbols must be mixed with the side-information symbols ($2$-sums) downloaded from the other server that do not contain desired message symbols.  This gives us the following query structure.
\begin{eqnarray*}
\begin{array}{c}
\theta=1\\
\begin{array}{|c|c|c|c|c|}\hline
\mbox{\small Server 1}&\mbox{\small Server 2}\\\hline
 a_1,  b_1,  c_1,  d_1& a_2,  b_2,  c_2,  d_2\\
 a_3 +  b_2 &  a_6 + b_1 \\
 a_4 + c_2 &  a_7 + c_1 \\
 a_5 + d_2 &  a_8 + d_1 \\
 b_4 + c_3 &  b_7 + c_6\\
 b_5 + d_3 &  b_8 + d_6 \\
 c_5 + d_4 &  c_8 +  d_7 \\
 a_9 +  b_7 +  c_6 &   a_{12} + b_4 +  c_3\\
 a_{10} + b_8 +  d_6 &   a_{13} +  b_5 +  d_3\\
 a_{11} +  c_8 +  d_7 &  a_{14} + c_5 +  d_4 \\
 b_* + c_* +  d_* &   b_* + c_* +  d_* \\
 a_{15} + b_*+ c_* +  d_* &  a_{16} +  b_* +  c_* +  d_*\\
\hline
\end{array}
\end{array}
\end{eqnarray*}
Now, again there is a special property that is evident to each server based on the $3$-sums that contain symbols from message $a$. Suppose we choose any two messages, one of which is $a$. For example, suppose we choose $a,b$ and consider Server $1$. Then there are $2$ instances of $3$-sums that contain $a,b$, namely, $a_9+b_7+c_6$ and $a_{10}+b_8+d_6$. Note that the third symbol in each case has the same index ($6$ in this case). The same is true if for example, we choose $a,c$ or $a,d$ instead. The two $3$-sums that contain $a,c$ are $a_9+b_7+c_6$ and $a_{11}+c_8+d_7$, and in each case the third symbol has the \emph{same} index ($7$ in this case). The two $3$-sums that contain $a,d$ are $a_{10}+b_8+d_6$ and $a_{11}+c_8+d_7$, and in each case the third symbol has the \emph{same} index ($8$ in this case). Again, because we do not wish to expose $a$ as the desired message, the same property must be true for all messages. This observation fixes the indices of the remaining $3$-sum, $b_*+c_*+d_*$ as follows. The index of $d$ in this term must be $9$ because the two $3$-sums that contain $b,c$ must have the same index for the third symbol, and according to $a_9+b_7+c_6$ this index must be $9$. Similarly, the index of $c$ in $b_*+c_*+d_*$ must be $10$ because the two $3$-sums that contain $b,d$ must have the same index for the third term, and according to $a_{10}+b_8+d_6$ it has to be $10$. The index of $b$ in $b_*+c_*+d_*$ is similarly determined by the term $a_{11}+c_8+d_7$ to be $11$. Thus, the query $b_*+c_*+d_*$ from Server $1$ must be $b_{11}+c_{10}+d_9$. Similarly, the query $b_*+c_*+d_*$ from Server $2$ must be $b_{14}+c_{13}+d_{12}$. 

The last step is again a side-information exploitation step, for which index assignment is trivial (new desired symbol must be combined with the $3$-sums queried from the other server that do not contain the desired symbol). Thus, the index assignment is complete, giving us the queries for $\PIRB$.
\begin{eqnarray*}
\begin{array}{c}
\theta=1\\
\begin{array}{|c|c|c|c|c|}\hline
\mbox{\small Server 1}&\mbox{\small Server 2}\\\hline
 a_1,  b_1,  c_1,  d_1& a_2,  b_2,  c_2,  d_2\\
 a_3 +  b_2 &  a_6 + b_1 \\
 a_4 + c_2 &  a_7 + c_1 \\
 a_5 + d_2 &  a_8 + d_1 \\
 b_4 + c_3 &  b_7 + c_6\\
 b_5 + d_3 &  b_8 + d_6 \\
 c_5 + d_4 &  c_8 +  d_7 \\
 a_9 +  b_7 +  c_6 &   a_{12} + b_4 +  c_3\\
 a_{10} + b_8 +  d_6 &   a_{13} +  b_5 +  d_3\\
 a_{11} +  c_8 +  d_7 &  a_{14} + c_5 +  d_4 \\
 b_{11} + c_{10} +  d_9 &   b_{14} + c_{13} +  d_{12} \\
 a_{15} + b_{14}+ c_{13} +  d_{12} &  a_{16} +  b_{11} +  c_{10} +  d_9\\
\hline
\end{array}
\end{array}
\end{eqnarray*}
For the sake of comparison, here are the queries generated with $\PIRB$ when $\theta=3$, i.e., when message $W_3$ (symbols $c$) is desired.
\begin{eqnarray*}
\begin{array}{c}
\theta=3\\
\begin{array}{|c|c|c|c|c|}\hline
\mbox{\small Server 1}&\mbox{\small Server 2}\\\hline
 a_1,  b_1,  c_1,  d_1& a_2,  b_2,  c_2,  d_2\\
 c_3 + a_2 &  c_6 + a_1 \\
 c_4 + b_2 &  c_7 + b_1 \\
 c_5 + d_2 &  c_8 + d_1 \\
 a_4 + b_3 &  a_7 + b_6\\
 a_5 + d_3 &  a_8 + d_6 \\
 b_5 + d_4 &  b_8 +  d_7 \\
 c_9 +  a_7 +  b_6 &   c_{12} + a_4 +  b_3\\
 c_{10} + a_8 +  d_6 &   c_{13} +  a_5 +  d_3\\
 c_{11} +  b_8 +  d_7 &  c_{14} + b_5 +  d_4 \\
 a_{11} + b_{10} +  d_9 &  a_{14} + b_{13} +  d_{12} \\
 c_{15} + a_{14}+ b_{13} +  d_{12} &  c_{16} +  a_{11} +  b_{10} +  d_9\\
\hline
\end{array}
\end{array}
\end{eqnarray*}
To see why the queries for $\theta=1$ are indistinguishable from the queries for $\theta=3$ under $\PIRB$, say from the perspective of Server $1$, note that the former is mapped to latter under the permutation on $[1:L]$ that maps
\begin{eqnarray}
&& \lefteqn{(1, 2,3,4, 5, 6, 7, 8, 9, 10, 11, 12, 13, 14, 15, 16)}\nonumber\\
&\longrightarrow&(1, 3,4,2, 5, 9, 6, 10, 7, 11, 8, 12, 15, 13, 14, 16)\nonumber
\end{eqnarray}
The permutation $\pi$ is chosen privately and uniformly by the user independent of $\theta$, so both queries are equally likely whether $\theta=1$ or $\theta=3$.

\subsubsection{\bf Arbitrary $K, M, N$}\label{kmn}

The extension to arbitrary $M, N$ is formally presented\footnote{Both $\PIRB$ and $\PC$ may be viewed as PIR schemes for $N$ servers with $M$ independent messages, so that $K$ is not directly needed for the query construction. Linear dependencies, if they are present, make some of the queries redundant, and allow a reduction in the number of downloaded symbols. $K$ only matters because it determines the number of redundant queries. The specific linear combinations involved in the $M$ functions are also not needed for the query construction. Thus the query construction has an intriguing `universal' character that exploits linear dependencies while remaining oblivious to the specifics of those dependencies.} in the query generation algorithm, {\bf Q-Gen}, that appears at the end of this section. Let us summarize the main ideas behind the generalization with the aid of the illustration in Figure \ref{fig:qtree} for $M=4, N=3$.
{\footnotesize
\begin{eqnarray*}
\hspace{-0.75cm}
\begin{array}{c}
\theta=1\\
\begin{array}{|c|c|c|c|c|}\hline
 B&\mbox{ Server } $1$ & \mbox{ Server } $2$& \mbox{ Server } $3$\\ \hline
1&a_1,  b_1, c_1, d_1& \begin{array}{ll} \cellcolor{red!10} a_2,~\tikzmark{x}& \tikzmark{y}~~\cellcolor{blue!10}b_2, c_2, d_2 \end{array}\tikzmark{a} & a_3, b_3, c_3, d_3\\ \hline
2&Q(1,2, \mathcal{M}): 
 \begin{array}{c}
 a_4+b_2 \\
 a_5+c_2 \\
 a_6 +d_2
 \end{array}~\tikzmark{k}
 & Q(2,1, \mathcal{M}):
  \begin{array}{c}
 a_{10}+b_1 \\
 a_{11}+c_1 \\
 a_{12} +d_1
 \end{array}
 & Q(3,1, \mathcal{M}):
   \begin{array}{c}
 a_{16}+b_1 \\
 a_{17}+c_1 \\
 a_{18} +d_1
 \end{array}
 \\  \cdashline{2-5}
 &Q(1,2, \mathcal{I}):
 \begin{array}{c}
 b_5 + c_4 \\
 b_6 + d_4 \\
 c_6 + d_5
 \end{array}~\tikzmark{l}
 &Q(2, 1, \mathcal{I}):
 \begin{array}{c}
 b_{11} + c_{10} \\
 b_{12} + d_{10} \\
 c_{12} + d_{11}
 \end{array}  
 &Q(3, 1, \mathcal{I}):
  \begin{array}{c}
 b_{17} + c_{16} \\
 b_{18} + d_{16} \\
 c_{18} + d_{17}
 \end{array} 
\\ \cline{2-5}
 &Q(1,3, \mathcal{M}):
 \begin{array}{c}
 a_7 + b_3 \\
 a_8 + c_3 \\
 a_9 + d_3
 \end{array} 
 &Q(2,3, \mathcal{M}):
  \begin{array}{c}
 a_{13} + b_3 \\
 a_{14} + c_3 \\
 a_{15} + d_3 
 \end{array} 
 &\tikzmark{b}~\cellcolor{red!10}Q(3,2, \mathcal{M}):
   \begin{array}{c}
 a_{19} + b_2 \\
 a_{20} + c_2 \\
 a_{21} + d_2 
 \end{array} 
 \\ \cdashline{2-5}
  &Q(1,3, \mathcal{I}):
 \begin{array}{c}
 b_8 + c_7 \\
 b_9 + d_7 \\
 c_9 + d_8
 \end{array}  
 &Q(2,3, \mathcal{I}):
  \begin{array}{c}
 b_{14} + c_{13} \\
 b_{15} + d_{13} \\
 c_{15} + d_{14}
 \end{array}  
 &\tikzmark{c}~\cellcolor{blue!10}Q(3,2, \mathcal{I}):
   \begin{array}{c}
 b_{20} + c_{19} \\
 b_{21} + d_{19} \\
 c_{21} + d_{20}
 \end{array}  
 \\  \hline
 3& Q(1,2, 1, \mathcal{M}): 
 \begin{array}{c}
 a_{22}+  b_{11} + c_{10} \\
 a_{23} + b_{12} + d_{10} \\
 a_{24} + c_{12} + d_{11}
 \end{array}
 &Q(2,1, 2, \mathcal{M}):
  \begin{array}{c}
 a_{34}+  b_{5} + c_{4} \\
 a_{35} + b_{6} + d_{4} \\
 a_{36} + c_{6} + d_{5}
 \end{array}
 &Q(3,1, 2, \mathcal{M}):
   \begin{array}{c}
 a_{46}+  b_{5} + c_{4} \\
 a_{47} + b_{6} + d_{4} \\
 a_{48} + c_{6} + d_{5}
 \end{array}
  \\ \cdashline{2-5}
  & Q(1,2, 1, \mathcal{I}): b_{24} + c_{23} + d_{22}
   &Q(2,1, 2, \mathcal{I}): b_{36} + c_{35} + d_{34}
 &Q(3,1, 2, \mathcal{I}): b_{48} + c_{47} + d_{46}
    \\ \cline{2-5}
  &  Q(1,2, 3, \mathcal{M}): 
   \begin{array}{c}
 a_{25}+  b_{14} + c_{13} \\
 a_{26} + b_{15} + d_{13} \\
 a_{27} + c_{15} + d_{14}
 \end{array}
    &  Q(2,1, 3, \mathcal{M}):
 \begin{array}{c}
 a_{37}+  b_{8} + c_{7} \\
 a_{38} + b_{9} + d_{7} \\
 a_{39} + c_{9} + d_{8}
 \end{array}
  &  Q(3,1, 3, \mathcal{M}):
   \begin{array}{c}
 a_{49}+  b_{8} + c_{7} \\
 a_{50} + b_{9} + d_{7} \\
 a_{51} + c_{9} + d_{8}
 \end{array}
      \\ \cdashline{2-5}
   &  Q(1,2, 3, \mathcal{I}): b_{27} + c_{26} + d_{25}
       &  Q(2,1, 3, \mathcal{I}): b_{39} + c_{38} + d_{37}
  &  Q(3,1, 3, \mathcal{I}): b_{51} + c_{50} + d_{49}
      \\ \cline{2-5}
    &  Q(1,3, 1, \mathcal{M}):
       \begin{array}{c}
 a_{28}+  b_{17} + c_{16} \\
 a_{29} + b_{18} + d_{16} \\
 a_{30} + c_{18} + d_{17}
 \end{array}
        &  Q(2,3, 1, \mathcal{M}):
               \begin{array}{c}
 a_{40}+  b_{17} + c_{16} \\
 a_{41} + b_{18} + d_{16} \\
 a_{42} + c_{18} + d_{17}
 \end{array}
    &  Q(3,2, 1, \mathcal{M}):
                   \begin{array}{c}
 a_{52}+  b_{11} + c_{10} \\
 a_{53} + b_{12} + d_{10} \\
 a_{54} + c_{12} + d_{11}
 \end{array}
      \\ \cdashline{2-5}
   &  Q(1,3, 1, \mathcal{I}): b_{30} + c_{29} + d_{28}
      &  Q(2,3, 1, \mathcal{I}): b_{42} + c_{41} + d_{40}
   &  Q(3,2, 1, \mathcal{I}): b_{54} + c_{53} + d_{52}
      \\ \cline{2-5}
     &  \cellcolor{red!10}Q(1,3, 2, \mathcal{M}):
            \begin{array}{c}
 a_{31}+  b_{20} + c_{19} \\
 a_{32} + b_{21} + d_{19} \\
 a_{33} + c_{21} + d_{20}
 \end{array}~\tikzmark{d}
      &  Q(2,3, 2, \mathcal{M}):
            \begin{array}{c}
 a_{43}+  b_{20} + c_{19} \\
 a_{44} + b_{21} + d_{19} \\
 a_{45} + c_{21} + d_{20}
 \end{array} \tikzmark{g}
      &  Q(3,2, 3, \mathcal{M}):
            \begin{array}{c}
 a_{55}+  b_{14} + c_{13} \\
 a_{56} + b_{15} + d_{13} \\
 a_{57} + c_{15} + d_{14}
 \end{array}
      \\ \cdashline{2-5}
   &  \cellcolor{blue!10} Q(1,3, 2, \mathcal{I}): b_{33} + c_{32} + d_{31}~~\tikzmark{e}
      &  Q(2,3, 2, \mathcal{I}): b_{45} + c_{44} + d_{43} ~~\tikzmark{h}
   &  Q(3,2, 3, \mathcal{I}): b_{57} + c_{56} + d_{55}
   \\ \hline
4&   Q(1,2,1, 2, \mathcal{M}): a_{58} + b_{36} + c_{35} + d_{34}
&   Q(2,1,2, 1, \mathcal{M}): a_{66} + b_{24} + c_{23} + d_{22}
&   Q(3,1,2, 1, \mathcal{M}): a_{74} + b_{24} + c_{23} + d_{22}
\\ \cline{2-5}
&   Q(1,2,1, 3, \mathcal{M}): a_{59} + b_{39} + c_{38} + d_{37}
&   Q(2,1,2, 3, \mathcal{M}): a_{67} + b_{27} + c_{26} + d_{25}
&   Q(3,1,2, 3, \mathcal{M}): a_{75} + b_{27} + c_{26} + d_{25}
\\ \cline{2-5}
&   Q(1,2,3, 1, \mathcal{M}): a_{60} + b_{42} + c_{41} + d_{40}
&   Q(2,1,3, 1, \mathcal{M}): a_{68} + b_{30} + c_{29} + d_{28}
&   Q(3,1,3, 1, \mathcal{M}): a_{76} + b_{30} + c_{29} + d_{28}
\\ \cline{2-5}
&   Q(1,2,3, 2, \mathcal{M}): a_{61} + b_{45} + c_{44} + d_{43}
&   \tikzmark{f}~\cellcolor{red!10}Q(2,1,3, 2, \mathcal{M}): a_{69} + b_{33} + c_{32} + d_{31}
&  \tikzmark{j}~ Q(3,1,3, 2, \mathcal{M}): a_{77} + b_{33} + c_{32} + d_{31}
\\ \cline{2-5}
&   Q(1,3,1, 2, \mathcal{M}): a_{62} + b_{48} + c_{47} + d_{46}
&   Q(2,3,1, 2, \mathcal{M}): a_{70} + b_{48} + c_{47} + d_{46}
&   Q(3,2,1, 2, \mathcal{M}): a_{78} + b_{36} + c_{35} + d_{34}
\\ \cline{2-5}
&   Q(1,3,1, 3, \mathcal{M}): a_{63} + b_{51} + c_{50} + d_{49}
&   Q(2,3,1, 3, \mathcal{M}): a_{71} + b_{51} + c_{50} + d_{49}
&   Q(3,2,1, 3, \mathcal{M}): a_{79} + b_{39} + c_{38} + d_{37}
\\ \cline{2-5}
&   Q(1,3,2, 1, \mathcal{M}): a_{64} + b_{54} + c_{53} + d_{52}
&   Q(2,3,2, 1, \mathcal{M}): a_{72} + b_{54} + c_{53} + d_{52}
&   Q(3,2,3, 1, \mathcal{M}): a_{80} + b_{42} + c_{41} + d_{40}
\\ \cline{2-5}
&   Q(1,3,2, 3, \mathcal{M}): a_{65} + b_{57} + c_{56} + d_{55}
&   Q(2,3,2, 3, \mathcal{M}): a_{73} + b_{57} + c_{56} + d_{55}
&  \tikzmark{i}~ Q(3,2,3, 2, \mathcal{M}): a_{81} + b_{45} + c_{44} + d_{43}
\\ \hline
\end{array}
\end{array}
\end{eqnarray*}
}
\begin{tikzpicture}[overlay, remember picture]
    \draw [blue, very thick, ->] ({pic cs:x})  to ({pic cs:y});
    \draw [red, very thick, ->] ({pic cs:a})  to ({pic cs:b});
    \draw [blue, very thick, ->] ({pic cs:b}) [bend right] to ({pic cs:c});
        \draw [red,very thick, ->] ({pic cs:c})  to ({pic cs:d});
                \draw [red, ->] ({pic cs:c})  to ({pic cs:g});
            \draw [blue,very thick, ->] ({pic cs:d}) [bend left] to ({pic cs:e});
                        \draw [blue, ->] ({pic cs:g}) [bend left] to ({pic cs:h});
                    \draw [red, very thick, ->] ({pic cs:e})  to ({pic cs:f});
                                        \draw [red,  ->] ({pic cs:e})  to ({pic cs:j});
                                        \draw [red,  ->] ([yshift=-.75pt]{pic cs:y})  to ({pic cs:k});
                        \draw [blue, ->] ({pic cs:k}) [bend left] to ({pic cs:l});
  \end{tikzpicture}
\begingroup
\captionof{figure}{\small Query generation tree according to $\PIRB$ for $M=4$ messages and $N=3$ servers. Red arrows indicate the use of the {\bf Exploit-SI} algorithm, and blue arrows indicate the use of the {\bf M-Sym} algorithm. Note that the symbol index assignments in any $\mathcal{I}$ partition are uniquely determined by the indices in the corresponding $\mathcal{M}$ partition.}\label{fig:qtree}
\endgroup

\bigskip
 The construction of queries for arbitrary $N$ servers is essentially a tree-like expansion of the $N=2$ construction. Therefore, the main insights all come from the $N=2$ setting. In fact, the index assignment process for $K$ messages is comprised of localized operations within the sets of queries that form the vertices of this tree, that operate exactly as in the $N=2$ setting. Let us use the tree terminology to explain the query construction for arbitrary $K,M,N$.

The root node (not shown because it carries no information) branches into $N$ vertices at depth $1$. These vertices, denoted $Q(n_1), n_1\in[1:N]$, represent the first set of queries from each server. For our example, $Q(n_1)=(a_{n_1}, b_{n_1}, c_{n_1}, d_{n_1})$. The queries associated with a vertex are internally partitioned into two parts. Queries that include a desired message symbol have the identifier $\mathcal{M}$, and queries that do not include any desired message symbol have the identifier $\mathcal{I}$. For our example we assume $\theta=1$, so that the $a_{n_1}$ symbols are the desired message symbols. Thus, $Q(n_1,\mathcal{M})=a_{n_1}$ and $Q(n_1,\mathcal{I})=(b_{n_1}, c_{n_1}, d_{n_1})$. 

Each level $1$ vertex, $Q(n_1), n_1\in[1:N]$, branches into $N-1$ vertices\footnote{A query vertex at level $m$ refers to the set of queries $Q(n_m, \cdots, n_1) = Q(n_m, \cdots, n_1, \mathcal{M}) \cup Q(n_m, \cdots, n_1, \mathcal{I}).$}, $Q(n_2, n_1), n_2\in[1:N], n_2\neq n_1$, to produce level $2$ of the tree. The query vertex $Q(n_2, n_1)$ is assigned to Server $n_2$. Thus, level $1$ vertices at Server $n_1$ generate level $2$ vertices associated with every server other than Server $n_1$. As a result each Server $n_2$, $n_2\in[1:N]$, has $N-1$ level $2$ query vertices, denoted $Q(n_2, n_1)$ for all $n_1\in[1:N], n_1\neq n_2$. Level $2$ query vertices are all comprised of $2$-sums, i.e., sums of two symbols, and are internally partitioned into $\mathcal{M}$ and $\mathcal{I}$ based on whether or not they contain desired message symbols. The queries in $Q(n_2, n_1, \mathcal{M})$ are generated by exploiting the side-information (cf. the {\bf Exploit-SI} algorithm \cite{Sun_Jafar_PIRL}) contained in the level $1$ queries $Q(n_1, \mathcal{I})$, i.e., these queries are generated by adding a new desired message symbol to each of the symbols in $Q(n_1,\mathcal{I})$. Thus, the query set $Q(n_2, n_1, \mathcal{M})$ contains $M-1$ elements. For our example, these $M-1=3$ elements are $Q(n_2, n_1, \mathcal{M})=\{a_i+b_{n_1}, a_j+c_{n_1}, a_k+d_{n_1}\}$, where $i,j,k$ are new symbol indices that have not appeared in any queries so far. Next, the queries in $Q(n_2, n_1, \mathcal{I})$ are generated to enforce message symmetry (cf. the {\bf M-Sym} algorithm \cite{Sun_Jafar_PIRL}), and contain a $2$-sum of every type that does not include the desired message, for a total of $\binom{M-1}{2}$ elements. For our example, these $\binom{3}{2}=3$ queries are $b_*+c_*, b_*+d_*, c_*+d_*$.  The symbol indices `*' are assigned based on the query set $Q(n_2, n_1, \mathcal{M})$ as described in our previous example. Since $Q(n_2, n_1, \mathcal{M})=\{a_i+b_{n_1}, a_j+c_{n_1}, a_k+d_{n_1}\}$ the index assignment produces $Q(n_2, n_1, \mathcal{I})=\{b_j+c_i, b_k+d_i, c_k+d_j\}$.

The query tree grows similarly to a total of $M$ levels.  A level $m$ query vertex assigned to Server $n_m$, $n_{m}\in[1:N]$, is denoted as $Q(n_m, n_{m-1}, \cdots, n_1)$ and is comprised of $m$-sums that include desired message symbols, denoted $Q(n_m, n_{m-1}, \cdots, n_1, \mathcal{M})$, and $m$-sums that do not include desired message symbols, denoted $Q(n_m, n_{m-1}, \cdots, n_1, \mathcal{I})$. The queries in $Q(n_m, n_{m-1}, \cdots, n_1, \mathcal{M})$ are $m$-sums generated by adding a new desired message symbol to each query contained in $Q(n_{m-1}, \cdots, n_1, \mathcal{I})$. This is  formalized in the {\bf Exploit-SI} algorithm. The queries in $Q(n_m, n_{m-1}, \cdots, n_1, \mathcal{I})$ are generated by the {\bf M-Sym} algorithm to force message symmetry, and contain an $m$-sum of every type that does not include the desired message, for a total of $\binom{M-1}{m}$ elements.\footnote{If $m=M$, then $Q(n_m, n_{m-1}, \cdots, n_1, \mathcal{I})$ is the empty set.} The index assignment for these queries takes place as follows. Consider a query $q\in Q(n_m, n_{m-1}, \cdots, n_1, \mathcal{I})$,  $q=u_{i_1}(*)+u_{i_2}(*)+\cdots+u_{i_m}(*)$,  where $*$ symbols represent indices that need to be assigned. Note that since this query is in the $\mathcal{I}$ partition, $\theta\notin \{i_1, i_2, \cdots, i_m\}$. The index $*$ for $u_{i_l}(*)$, $l\in[1:m]$, comes from the $m$-sum query in $Q(n_m, n_{m-1}, \cdots, n_1, \mathcal{M})$ that contains symbols from $u_{i_1}, u_{i_2}, \cdots, u_{i_{l-1}}, u_\theta, u_{i_{l+1}}, \cdots  u_{i_m}$. If the symbol index for $u_\theta$ in this query is $j_l$, i.e., the query contains $u_\theta(j_l)$ then the  index $j_l$ is assigned to $u_{i_l}$. In this way, the {\bf M-Sym} algorithm assigns all indices to generate the query $q=u_{i_1}(j_1)+u_{i_2}(j_2)+\cdots+u_{i_m}(j_m)$. This completes the description of $\PIRB$.

The following observations follow immediately from the query construction described above.
\begin{enumerate}
\item $|Q(n_m, n_{m-1}, \cdots, n_1, \mathcal{I})|=\binom{M-1}{m}$
\item $|Q(n_m, n_{m-1}, \cdots, n_1, \mathcal{M})|=|Q(n_{m-1}, \cdots, n_1, \mathcal{I})|=\binom{M-1}{m-1}$
\item The number of level $m$ query vertices $Q(n_m, n_{m-1}, \cdots, n_1)$ assigned to Server $i$, (such that $n_m=i$), is $(N-1)^{m-1}$. This is  because there are $N-1$ valid values for $n_{m-1}$ that are not equal to $n_m=i$, there are $N-1$ values for $n_{m-2}$ that are not equal to $n_{m-1}$, and so on.
\item The total number of queries assigned to Server $i$ is 
$\sum_{m=1}^{M}(N-1)^{m-1}\left(\binom{M-1}{m}+\binom{M-1}{m-1}\right)$.
\item If $Q$ and $Q'$ are two query vertices assigned to the same server, then the symbol indices that appear in $Q$ are distinct from the symbol indices that appear in $Q'$.
\end{enumerate}

The proof of privacy for $\PIRB$ is similar to that for $\PIRA$ in \cite{Sun_Jafar_PIR}. We note that once the labels $\mathcal{M}, \mathcal{I}$ are suppressed, and the queries sorted in lexicographic order, the structure of the queries from any individual server is fixed regardless of the desired message index $\theta$. For our $M=4, N=3$ example, this is illustrated in Figure \ref{fig:qstruct}.

Note that only distinct symbol indices are shown. All the remaining indices can be inferred uniquely from the ones shown based on the index assignment rule. Thus, the particular query realization (depending on $\theta$) to Server $n$, $n\in[1:N]$, depends only on the realization of these distinct indices. However, the indices depend on the permutation $\pi$ which is  chosen uniformly and privately by the user. Thus, all distinct choices for these indices are equally likely, regardless of $\theta$, and the scheme is private.

The correctness of $\PIRB$ follows directly from the correctness of $\PIRA$. By the same token, if the  messages are independent then $\PIRA$ and $\PIRB$ have the same rate. 
Thus, the index assignment process produces a new PIR scheme, $\PIRB$, that for independent messages, is equally efficient as $\PIRA$  in terms of download, i.e., $\PIRB$ is capacity achieving for independent messages. 
However, depending upon the form of the message dependencies, it turns out that the `sums' may not be sufficient and more sophisticated mixing of message symbols may be required.  For the linear dependencies that we consider in this paper, we will need sign assignments, that are explained next.

{\small
\begin{eqnarray*}
\begin{array}{|c|c|c|c|c|}\hline
 B&\mbox{ Server } n \\ \hline
1&Q(1):a_{i_1},  b, c, d\\ \hline
2&Q(1,2): 
 \begin{array}{c}
 a_{j_{2,(1,2)}}+b_{i_{2,(1,2)}} \\
 a_{k_{2,(1,2)}}+c \\
 a_{l_{2,(1,2)}} +d\\
 b + c \\
 b + d \\
 c + d
 \end{array}\\ \cline{2-5}
 &Q(1,3):
 \begin{array}{c}
 a_{j_{2,(1,3)}} + b_{i_{2,(1,3)}} \\
 a_{k_{2,(1,3)}} + c \\
 a_{l_{2,(1,3)}} + d \\
 b + c \\
 b + d \\
 c + d
 \end{array}  \\  \hline
 3& Q(1,2, 1): 
 \begin{array}{c}
 a_{l_{3,(1,2,1)}}+  b_{j_{3,(1,2,1)}} + c_{i_{3,(1,2,1)}} \\
 a_{s_{3,(1,2,1)}} + b_{k_{3,(1,2,1)}} + d \\
 a_{t_{3,(1,2,1)}} + c + d \\
 b + c + d
 \end{array} \\ \cline{2-5}
  &  Q(1,2, 3): 
   \begin{array}{c}
 a_{l_{3,(1,2,3)}}+  b_{j_{3,(1,2,3)}} + c_{i_{3,(1,2,3)}} \\
 a_{s_{3,(1,2,3)}} + b_{k_{3,(1,2,3)}} + d \\
 a_{t_{3,(1,2,3)}} + c + d \\
 b + c + d
 \end{array}     \\ \cline{2-5}
    &  Q(1,3, 1):
       \begin{array}{c}
 a_{l_{3,(1,3,1)}}+  b_{j_{3,(1,3,1)}} + c_{i_{3,(1,3,1)}} \\
 a_{s_{3,(1,3,1)}} + b_{k_{3,(1,3,1)}} + d \\
 a_{t_{3,(1,3,1)}} + c + d \\
 b + c + d
 \end{array}      \\ \cline{2-5}
     &  Q(1,3, 2):
            \begin{array}{c}
 a_{l_{3,(1,3,2)}}+  b_{j_{3,(1,3,2)}} + c_{i_{3,(1,3,2)}} \\
 a_{s_{3,(1,3,2)}} + b_{k_{3,(1,3,2)}} + d \\
 a_{t_{3,(1,3,2)}} + c + d \\
 b + c + d
 \end{array}\\ \hline
4&   Q(1,2,1, 2): a_{l_{4,(1,2,1,2)}} + b_{k_{4,(1,2,1,2)}} + c_{j_{4,(1,2,1,2)}} + d_{i_{4,(1,2,1,2)}} \\ \cline{2-5}
&   Q(1,2,1, 3): a_{l_{4,(1,2,1,3)}} + b_{k_{4,(1,2,1,3)}} + c_{j_{4,(1,2,1,3)}} + d_{i_{4,(1,2,1,3)}}\\ \cline{2-5}
&   Q(1,2,3, 1): a_{l_{4,(1,2,3,1)}} + b_{k_{4,(1,2,3,1)}} + c_{j_{4,(1,2,3,1)}} + d_{i_{4,(1,2,3,1)}}\\ \cline{2-5}
&   Q(1,2,3, 2): a_{l_{4,(1,2,3,2)}} + b_{k_{4,(1,2,3,2)}} + c_{j_{4,(1,2,3,2)}} + d_{i_{4,(1,2,3,2)}}\\ \cline{2-5}
&   Q(1,3,1, 2): a_{l_{4,(1,3,1,2)}} + b_{k_{4,(1,3,1,2)}} + c_{j_{4,(1,3,1,2)}} + d_{i_{4,(1,3,1,2)}}\\ \cline{2-5}
&   Q(1,3,1, 3): a_{l_{4,(1,3,1,3)}} + b_{k_{4,(1,3,1,3)}} + c_{j_{4,(1,3,1,3)}} + d_{i_{4,(1,3,1,3)}}\\ \cline{2-5}
&   Q(1,3,2, 1): a_{l_{4,(1,3,2,1)}} + b_{k_{4,(1,3,2,1)}} + c_{j_{4,(1,3,2,1)}} + d_{i_{4,(1,3,2,1)}}\\ \cline{2-5}
&   Q(1,3,2, 3): a_{l_{4,(1,3,2,3)}} + b_{k_{4,(1,3,2,3)}} + c_{j_{4,(1,3,2,3)}} + d_{i_{4,(1,3,2,3)}}\\ \hline
\end{array}
\end{eqnarray*}
}
\begingroup
\captionof{figure}{\small Structure of queries generated by $\PIRB$ when $M=4$ and $N=3$.}\label{fig:qstruct}
\endgroup

\bigskip

\subsection{Sign Assignment: $\PC$}
In this section, we present the sign assignment procedure that produces the private computation scheme $\PC$ from $\PIRB$ for arbitrary $K, M, N$. We will use Example $A$ to illustrate its steps. The sign assignment procedure depends on $\theta$. Let us choose $\theta=3$ to illustrate the process. Note that $\sigma_i$ are now generated uniformly and independently from $\{+1, -1\}$.

To explain the sign assignment, it is convenient to express each query in lexicographic order. For example, the query $u_{i_1}(j_1)+u_{i_2}(j_2)+\cdots+u_{i_m}(j_m)$ is in lexicographic order if $i_1<i_2<\cdots<i_m$ regardless of the values of the indices $j$. For our $M=4$ example, the query  $c_9+a_7+b_6$ is expressed as $a_7+b_6+c_9$ under lexicographic ordering. Note that the lexicographic order for the $M=4$ example is simply the ordering $a<b<c<d$ and the indices do not matter. The position of the  $c_*$ symbol within this lexicographic ordering of query $q$ will be denoted as $\Delta_{c}(q)$, i.e., for the query $q=a_7+b_6+c_9$, we have $\Delta_a(q)=1, \Delta_b(q)=2, \Delta_c(q)=3$ and $\Delta_d(q)=0$ where the $0$ value indicates that a symbol from that message is not present in the query.

Next, the queries are sorted in increasing order of blocks, $B$, so that the $m^{th}$ block $B=m$, contains only $m$-sums. Each block is partitioned into sub-blocks, $S$, such that all the queries $q$ in the same sub-block have the same value of $\Delta_{W_\theta}(q)$. The sub-blocks are sorted within a block  in \emph{descending} order of $\Delta_{W_\theta}(q)$ and numbered $S=1,2, \cdots$. With this sorting, the query structure  is represented as follows.
\begin{eqnarray*}
\begin{array}{c}
\theta=3\\
\begin{array}{|c|c|c|c|c|}\hline
B& S(\Delta_{c}) &\mbox{ Server } $1$ & \mbox{ Server } $2$\\ \hline
1&\cdots&c_1, a_1,  b_1, d_1& c_2, a_2, b_2, d_2\\ \hline
2 & 1(2)&a_2+c_3  & a_1+c_6\\
&1(2)& b_2+c_4 & b_1+c_7\\
&2(1)& c_5 + d_2 &c_8+d_1 \\
 &3(0)&a_4 + b_3 &a_7+b_6 \\
 &3(0)&a_5 + d_3 & a_8+d_6\\
 &3(0)&b_5 + d_4 &b_8+d_7\\ \hline
 3&1(3)&a_7+b_6 + c_9 & a_4+b_3+c_{12}\\
 &2(2)&a_8 +c_{10} +   d_6 &a_5+c_{13}+d_3\\
 &2(2)&b_8 +c_{11} +    d_7 &b_5+c_{14}+d_4\\
 &3(0)&a_{11} + b_{10} +  d_9  &a_{14}+b_{13}+d_{12}\\ \hline
 4&1(3)&a_{14}+ b_{13} +c_{15} +   d_{12}&a_{11}+b_{10}+c_{16}+d_9 \\
\hline
\end{array}
\end{array}
\end{eqnarray*}

The sign assignment algorithm for arbitrary $M$ is comprised of $4$ steps.

\noindent{\bf Algorithm: SignAssign}
\begin{enumerate}
\item[] ({\bf Step 1}) Consider queries for which $\Delta_{W_\theta}(q)=0$, i.e., queries that do not contain desired message symbols. The terms in these queries that occupy even positions (in lexicographic order within each query) are assigned the `$-$' sign. Thus, for example the query $q= a_{11}+b_{10}+d_9$ changes to $q\rightarrow q' = a_{11}-b_{10}+d_9$ after the sign assignment. Notice that the signs are alternating in the lexicographic ordering of symbols within the query. The sign assignments for the queries with $\Delta_{W_\theta}(q)=0$  are now settled. 
\item[]({\bf Step 2}) If a symbol is assigned a negative sign in Step $1$ then in Step $2$ it is assigned a negative sign everywhere it appears. Note that any undesired symbol that appears in the query from one server, appears exactly once within the query to each server.

For our $M=4$ example, at this point we have,
\begin{eqnarray*}
\begin{array}{c}
\theta=3\\
\begin{array}{|c|c|c|c|c|}\hline
B& S(\Delta_{c}) &\mbox{ Server } $1$ & \mbox{ Server } $2$\\ \hline
1&\cdots&c_1, a_1,  b_1, d_1& c_2, a_2, b_2, d_2\\ \hline
2 & 1(2)&a_2+c_3  & a_1+c_6\\
&1(2)& b_2+c_4 & b_1+c_7\\ \cline{2-4}
&2(1)& c_5 + d_2 &c_8+d_1 \\ \cline{2-4}
 &3(0)&a_4 - b_3 &a_7-b_6 \\
 &3(0)&a_5 - d_3 & a_8-d_6\\
 &3(0)&b_5 - d_4 &b_8-d_7\\ \hline
 3&1(3)&a_7-b_6 + c_9 & a_4-b_3+c_{12}\\ \cline{2-4}
 &2(2)&a_8 +c_{10} -   d_6 &a_5+c_{13}-d_3\\
 &2(2)&b_8 +c_{11} -    d_7 &b_5+c_{14}-d_4\\ \cline{2-4}
 &3(0)&a_{11} - b_{10} +  d_9  &a_{14}- b_{13}+d_{12}\\ \hline
 4&1(3)&a_{14} - b_{13} +c_{15} +   d_{12}&a_{11}- b_{10}+c_{16}+d_9 \\
\hline
\end{array}
\end{array}
\end{eqnarray*}

\item[]({\bf Step 3}) Every query such that $\Delta_{W_\theta}(q)> 0$, i.e., every query that contains a desired message symbol  is multiplied by $(-1)^{S+1(\theta\neq 1)}$, where $S$ is the sub-block index and $1(\theta\neq 1)$ is the indicator function that takes the value $1$ if $\theta\neq 1$ and $0$ if $\theta=1$.

\item[]({\bf Step 4})  Finally, in Step $4$, for each query $q$ that contains a desired symbol, i.e., $\Delta_{W_\theta}(q)>0$, the desired symbol is assigned the negative sign if it occupies an even numbered position, i.e., if $\Delta_{W_\theta}(q)$ is an even number, and a positive sign if it occupies an odd numbered position, i.e., if $\Delta_{W_\theta}(q)$ is an odd number.

\end{enumerate}
Following this procedure for our running example, we have the final form of the queries as follows.
\begin{eqnarray*}
\begin{array}{c}
\theta=3\\
\begin{array}{|c|c|c|c|c|}\hline
B& S(\Delta_{c}) &\mbox{ Server } $1$ & \mbox{ Server } $2$\\ \hline
1&\cdots&c_1, a_1,  b_1, d_1& c_2, a_2, b_2, d_2\\ \hline
2 & 1(2)&a_2-c_3  & a_1-c_6\\
&1(2)& b_2-c_4 & b_1-c_7\\ \cline{2-4}
&2(1)& c_5 - d_2 &c_8 - d_1 \\ \cline{2-4}
 &3(0)&a_4 - b_3 &a_7-b_6 \\
 &3(0)&a_5 - d_3 & a_8-d_6\\
 &3(0)&b_5 - d_4 &b_8-d_7\\ \hline
 3&1(3)&a_7-b_6 + c_9 & a_4-b_3+c_{12}\\ \cline{2-4}
 &2(2)&-a_8 -c_{10} +   d_6 & - a_5-c_{13}+d_3\\
 &2(2)& - b_8 -c_{11} +    d_7 & - b_5-c_{14}+d_4\\ \cline{2-4}
 &3(0)&a_{11} - b_{10} +  d_9  &a_{14}- b_{13}+d_{12}\\ \hline
 4&1(3)&a_{14} - b_{13} +c_{15} +   d_{12}&a_{11}- b_{10}+c_{16}+d_9 \\
\hline
\end{array}
\end{array}
\end{eqnarray*}
To complete the illustration for our $M=4$ example, let us also present the final queries for $\theta=1, 2, 4$.
\begin{eqnarray*}
\begin{array}{c}
\theta=1\\
\begin{array}{|c|c|c|c|c|}\hline
B& S(\Delta_{c}) &\mbox{ Server } $1$ & \mbox{ Server } $2$\\ \hline
1&\cdots&a_1,  b_1, c_1, d_1& a_2, b_2, c_2, d_2\\ \hline
2 & 1(1)&a_3-b_2  & a_6-b_1\\
&1(1)& a_4-c_2 & a_7-c_1\\ 
&1(1)& a_5 - d_2 &a_8 - d_1 \\ \cline{2-4}
 &2(0)&b_4 -c_3 &b_7-c_6 \\
 &2(0)&b_5 - d_3 & b_8-d_6\\
 &2(0)&c_5 - d_4 &c_8-d_7\\ \hline
 3&1(1)&a_9 - b_7 + c_6 & a_{12} - b_4 + c_{3}\\
 &1(1)& a_{10} -b_{8} +   d_6 & a_{13}-b_5+d_3\\
 &1(1)& a_{11}- c_8 +d_{7}  & a_{14}-c_5+d_4\\ \cline{2-4}
 &2(0)&b_{11} - c_{10} +  d_{9}  &b_{14} -  c_{13}+d_{12}\\ \hline
 4&1(1)&a_{15} - b_{14} +c_{13} -  d_{12}&a_{16}- b_{11}+c_{10}-d_{9} \\
\hline
\end{array}
\end{array}
\end{eqnarray*}
\begin{eqnarray*}
\begin{array}{c}
\theta=2\\
\begin{array}{|c|c|c|c|c|}\hline
B& S(\Delta_{b}) &\mbox{ Server } $1$ & \mbox{ Server } $2$\\ \hline
1&\cdots&b_1,  a_1, c_1, d_1& b_2, a_2, c_2, d_2\\ \hline
2 & 1(2)&a_2-b_3  & a_1-b_6\\ \cline{2-4}
&2(1)& b_4-c_2 & b_7-c_1\\ 
&2(1)& b_5 - d_2 &b_8 - d_1 \\ \cline{2-4}
 &3(0)&a_4 -c_3 &a_7-c_6 \\
 &3(0)&a_5 - d_3 & a_8-d_6\\
 &3(0)&c_5 - d_4 &c_8-d_7\\ \hline
 3&1(2)&a_7 - b_9 - c_6 & a_{4} - b_{12} -c_{3}\\
 &1(2)& a_{8} -b_{10} -   d_6 & a_{5}-b_{13}-d_3\\\cline{2-4}
 &2(1)& b_{11}- c_8 +d_{7}  & b_{14}-c_5+d_4\\ \cline{2-4}
 &3(0)&a_{11} - c_{10} +  d_{9}  &a_{14} -  c_{13}+d_{12}\\ \hline
 4&1(2)&a_{14} - b_{15} -c_{13} +  d_{12}&a_{11}- b_{16}-c_{10}+d_{9} \\
\hline
\end{array}
\end{array}
\end{eqnarray*}
\begin{eqnarray*}
\begin{array}{c}
\theta=4\\
\begin{array}{|c|c|c|c|c|}\hline
B& S(\Delta_{d}) &\mbox{ Server } $1$ & \mbox{ Server } $2$\\ \hline
1&\cdots&d_1,  a_1, b_1, c_1& d_2, a_2, b_2, c_2\\ \hline
2 & 1(2)&a_2-d_3  & a_1-d_6\\ 
&1(2)& b_2-d_4 & b_1-d_7\\ 
&1(2)& c_2 - d_5 &c_1 - d_8 \\ \cline{2-4}
 &2(0)&a_4 -b_3 &a_7-b_6 \\
 &2(0)&a_5 - c_3 & a_8-c_6\\
 &2(0)&b_5 - c_4 &b_8-c_7\\ \hline
 3&1(3)&a_7 - b_6 +d_9 & a_{4} - b_{3} + d_{12}\\
 &1(3)& a_{8} -c_{6} +   d_{10} & a_{5}-c_{3} + d_{13}\\
 &1(3)& b_{8}- c_7 +d_{11}  & b_{5}-c_4+d_{14}\\ \cline{2-4}
 &2(0)&a_{11} - b_{10} +  c_{9}  &a_{14} -  b_{13}+c_{12}\\ \hline
 4&1(4)&a_{14} - b_{13} +c_{12} -  d_{15}&a_{11}- b_{10}+c_{9}-d_{16} \\
\hline
\end{array}
\end{array}
\end{eqnarray*}

We include the full algorithm here for completeness. $Q({n},\text{`}\theta\text{'})$ denotes the queries for Server $n \in [1:N]$ when $W_\theta$ is desired.  For any ordered tuple $u$, let $\new(u)$ be a function that, starting with $u(1)$,  returns the ``next" element in $u$ each time it is called with the same tuple $u$ as its argument. 
\allowdisplaybreaks
\begin{algorithm}[H]
\caption{{\bf Q-Gen} Algorithm.}
\label{alg1}
\begin{algorithmic}[1]{}
\STATE{{\bf Input:} $\theta$}
\STATE{{\bf Output:} $Q({1},\text{`}\theta\text{'}), \cdots, Q(N,\text{`}\theta\text{'})$}
\STATE{Initialize: All query sets are initialized as null sets. Also initialize $\m{Block}= 1$;} 
\FOR
{$\m{DB}_1=1:N$} 
\STATE{
\begin{align*}
&Q(\m{DB}_1,\text{`}\theta\text{'},\m{Block},\mathcal{M}) \leftarrow\{u_\theta(\m{DB}_1)\}\\~
& Q(\m{DB}_1,\text{`}\theta\text{'},\m{Block},\mathcal{I}) \leftarrow\{u_1(\m{DB}_1),  \cdots, u_M(\m{DB}_1)\}/\{u_\theta(\m{DB}_1)\}
\end{align*}
}
\ENDFOR $\m{(DB}_1)$
\FOR
{$\m{Block}=2:M$} 
\FOR
{${\color{black}\m{DB}_{\mbox{\tiny Block}}}=1:N$} 
\FOR
{{\color{black}each $(\m{DB}_{\mbox{\tiny Block}-1}, \m{DB}_{\mbox{\tiny Block}-2}, \cdots, \m{DB}_1)$, where $\m{DB}_{\mbox{\tiny Block}-1} \neq \m{DB}_{\mbox{\tiny Block}}, \m{DB}_{\mbox{\tiny Block}-2} \neq \m{DB}_{\mbox{\tiny Block}-1}, \cdots, \m{DB}_1 \neq \m{DB}_2$}
} 
\STATE{
{\small \setlength{\mathindent}{0pt}
\begin{align*}
&Q(\m{DB}_{\mbox{\tiny Block}},\m{DB}_{\mbox{\tiny Block}-1}, \cdots, \m{DB}_1, \text{`}\theta\text{'},\m{Block},\mathcal{M})  \leftarrow \m{\textbf{\color{black} {\bf Exploit-SI}}}(Q(\m{DB}_{\mbox{\tiny Block}-1}, \m{DB}_{\mbox{\tiny Block}-2}, \cdots, \m{DB}_{1},\text{`}\theta\text{'},\m{Block}-1,\mathcal{I}))\\
&Q(\m{DB}_{\mbox{\tiny Block}},\m{DB}_{\mbox{\tiny Block}-1}, \cdots, \m{DB}_1, \text{`}\theta\text{'},\m{Block},\mathcal{I}) \leftarrow \m{\textbf{\color{black} {\bf M-Sym}}}(Q(\m{DB}_{\mbox{\tiny Block}},\m{DB}_{\mbox{\tiny Block}-1}, \cdots, \m{DB}_1, \text{`}\theta\text{'},\m{Block},\mathcal{M}))\\
\end{align*}
}
}
\ENDFOR ~$(\m{DB}_{\mbox{\tiny Block}-1}, \m{DB}_{\mbox{\tiny Block}-2}, \cdots, \m{DB}_{1})$
\ENDFOR $\m{(DB}_{\mbox{\tiny Block}})$
\ENDFOR \m{ (Block)}
\FOR
{$\m{DB}_{\mbox{\tiny Block}}=1:N$} 
\STATE {
\begin{align*}
& Q(\m{DB}_{\mbox{\tiny Block}},\text{`}\theta\text{'})\leftarrow\bigcup_{\m{\tiny Block}\in[1:M]} \bigcup_{\begin{subarray}{c}
\mbox{\scriptsize DB}_{\mbox{\tiny Block}-1} \neq \mbox{\scriptsize DB}_{\mbox{\tiny Block}}, \\
\cdots, \mbox{\scriptsize DB}_{1} \neq \mbox{\scriptsize DB}_{2}
\end{subarray}
}  \big(Q(\m{DB}_{\mbox{\tiny Block}}, \m{DB}_{\mbox{\tiny Block}-1}, \cdots, \m{DB}_{1}, \text{`}\theta\text{'},\m{Block},\mathcal{I})\cup \notag\\
&~~~~~~~~~~~~~~~~~~~~~~~~~~~~~~~~~~~~~~~~~~~~~~~~~~~~~~~~~~~Q(\m{DB}_{\mbox{\tiny Block}}, \m{DB}_{\mbox{\tiny Block}-1}, \cdots, \m{DB}_{1}, \text{`}\theta\text{'},\m{Block},\mathcal{M})\big) 
\end{align*}
\ENDFOR $\m{(DB}_{\mbox{\tiny Block}})$
}
\STATE{{\bf SignAssign}$\left( Q({1},\text{`}\theta\text{'}), \cdots, Q(N,\text{`}\theta\text{'}) \right)$}
\end{algorithmic}
\end{algorithm}

The sub-routines are as follows. $\theta, \m{Block}$ are assumed to be available to the sub-routines as global variables. $\mathcal{T}_m$ represents the set of all possible choices of $m$ distinct indices in $[1:M]$. $\overrightarrow{\mathcal{T}}$ indicates that the elements of $\mathcal{T}$ are to be accessed in the natural lexicographic increasing order.

\allowdisplaybreaks
\begin{algorithm}[H]
\caption{{\bf M-Sym} Algorithm.}
\label{msym}
\begin{algorithmic}[1]{}
\STATE{{\bf Input:} $Q= Q(\m{DB}_{\mbox{\tiny Block}},\m{DB}_{\mbox{\tiny Block}-1}, \cdots, \m{DB}_1, \text{`}\theta\text{'},\m{Block},\mathcal{M})$}
\STATE{{\bf Output:} $Q^*=Q(\m{DB}_{\mbox{\tiny Block}},\m{DB}_{\mbox{\tiny Block}-1}, \cdots, \m{DB}_1, \text{`}\theta\text{'},\m{Block},\mathcal{I})$}
\STATE{Initialize: $Q^* \leftarrow \emptyset$.} 
\FOR
{\textbf{each} $i_{[1:\mbox{\scriptsize Block}]}\in \overrightarrow{\mathcal{T}_{\mbox{\scriptsize Block}}}$,  $\theta \notin i_{[1:\mbox{\scriptsize Block}]} $}
\STATE{
\begin{align*}
Q^* \leftarrow &Q^* \cup \{u_{i_1}(j_1)+u_{i_2}(j_2)+\cdots+u_{i_{\m{\scriptsize Block}}}(j_{\m{\scriptsize Block}})\}\\
&\mbox{such that } \forall l\in[1:\m{Block}]\\
&\exists \m{ } u_\theta(j_l)+\sum_{r\in[1:\m{\scriptsize Block}], r\neq l}u_{i_r}(*)\in Q
\end{align*}
} 
\ENDFOR \m{ ($i_{[1:\mbox{\scriptsize Block}]}$)}
\end{algorithmic}
\end{algorithm}

\allowdisplaybreaks
\begin{algorithm}[H]{}
\caption{{\bf Exploit-SI} Algorithm.}
\label{exploit}
\begin{algorithmic}[1]{}
\STATE{{\bf Input:} $Q=Q(\m{DB}_{\mbox{\tiny Block}-1}, \m{DB}_{\mbox{\tiny Block}-2}, \cdots, \m{DB}_{1},\text{`}\theta\text{'},\m{Block}-1,\mathcal{I})$}
\STATE{{\bf Output:} $Q'=Q(\m{DB}_{\mbox{\tiny Block}},\m{DB}_{\mbox{\tiny Block}-1}, \cdots, \m{DB}_1, \text{`}\theta\text{'},\m{Block},\mathcal{M})$}
\STATE{Initialize: $Q' \leftarrow \emptyset$.} 
\FOR
{\textbf{each} $q \in \overrightarrow{Q}$
}
\STATE{$$Q' \leftarrow Q' \cup \{\new(u_{\theta}) + q  \}$$} \ENDFOR \m{ ($q$)} 
\end{algorithmic}
\end{algorithm}

This completes the description of the scheme $\PC$. The correctness of $\PC$ follows from that of $\PIRB$. Remarkably, if the messages are independent, then $\PC$ may be seen as another PIR scheme that achieves the same rate as $\PIRA$, $\PIRB$, i.e., all three are capacity achieving schemes. The proof of privacy of $\PC$ is deferred to Section \ref{sec:privacy_toy} for Example $A$ and to Section \ref{sec:privacy_general} for arbitrary $K, M, N$. 

The main advantage of $\PC$ is that for the dependent message setting of Theorem \ref{thm}, it  is the optimal private computation scheme. Its proof of optimality  is presented next.
\section{Proof of Optimality of $\PC$}
In this section, we show how $\PC$ achieves the capacity of private computation when the messages are dependent. The key idea is that the message dependencies combined with the special index and sign structure of $\PC$ create redundant queries, and by downloading generic combinations of the queries\footnote{Alternatively, random binning (Slepian-Wolf coding) may be used.} instead of downloading each query separately,  the download requirement is reduced without compromising on privacy. 

\subsection{Proof of Optimality for Example $A$}
To prove optimality, we need to show that the scheme achieves a rate that matches the capacity of private computation according to Theorem \ref{thm}. Specifically, let us prove that the rate achieved is $8/12 = 2/3$. For this, we will show that the user downloads only $12$ symbols from each server. Note that ostensibly there are $15$ symbols that are queried from each server. However, it turns out that based on the information available from the other server, $3$ of these symbols are redundant. Thus, $12$ generic combinations of these $15$ symbols are sufficient.

Let us see why this is the case for the queries from Server 1.  $c_1, d_1$ are clearly redundant symbols because according to (\ref{eq:r1}) they are functions of $a_1, b_1$.  So we need one more redundant symbol. Suppose $a$ is desired ($\theta = 1$). Then, consider the $2$-sum queries that do not involve the desired message, $a$. There are $3$ such queries. However, the key is that from any $2$ we can construct the $3^{rd}$. In this case from Server 1 we have: $b_4 - c_3, b_5 - d_3, c_5 - d_4$. But note that
\begin{eqnarray*}
&&v'_3(b_5- d_3)-v'_4(b_4- c_3)- (v_3v'_4-v_4v'_3) a_3 - v_4 a_4+ v_3 a_5 =( c_5-  d_4)
\end{eqnarray*}
Verify:
\begin{eqnarray*}
\mbox{LHS }&=&v'_3(b_5-  d_3)-v'_4( b_4-  c_3)-  (v_3v'_4-v_4v'_3) a_3 -  v_4 a_4+  v_3 a_5\\
&\overset{(\ref{eq:r1})}{=}&v'_3(b_5-  v_4a_3- v'_4b_3)-v'_4(b_4-  v_3a_3-  v'_3b_3) - (v_3v'_4-v_4v'_3)a_3-  v_4 a_4+  v_3 a_5\\
&\overset{}{=}& v_3 a_5+ v'_3 b_5-  v_4 a_4- v'_4 b_4\overset{(\ref{eq:r1})}{=}( c_5-  d_4) = \mbox{RHS}
\end{eqnarray*}
Since the user  knows $a_3, a_4, a_5$ due to the side information available from the other server,  out of these $3$ equations, $1$ is redundant. Thus, one more symbol is saved, giving us $12$ effective downloaded symbols, and the rate $8/12$ is achieved. Since this is also the outer bound, this scheme achieves capacity. It can similarly be verified for Example $A$ that the redundancy exists no matter which message is desired. 

As another example, suppose $c$ is desired ($\theta = 3$). 
Referring to the scheme, from Server 1, the three queries (that are $2$-sums) not involving $c$ are $a_4- b_3$, $a_5 - d_3$, $b_5- d_4$. But note that
\begin{eqnarray*}
&& (v_3v'_4 - v_4v'_3)(a_4-  b_3) - v_3 ( a_5 -  d_3) - v_4c_3 - v'_4 c_4+  c_5 =v'_3 ( b_5-  d_4)
\end{eqnarray*}
Verify
\begin{eqnarray*}
\mbox{LHS} &=& (v_3v'_4 - v_4v'_3)( a_4- b_3) - v_3( a_5 -  d_3)- v_4c_3- v'_4 c_4+  c_5\\
&\overset{(\ref{eq:r1})}{=}& (v_3v'_4 - v_4v'_3)( a_4-  b_3) - v_3( a_5 -  v_4 a_3 - v'_4 b_3)\\
&&~- v_4(v_3a_3 + v'_3b_3) -v'_4 (v_3a_4 + v'_3b_4)+  (v_3a_5 + v'_3b_5)\\
&\overset{}{=}&v'_3 (b_5- v_4 a_4-v'_4 b_4)\\
&\overset{(\ref{eq:r1})}{=}&v'_3 ( b_5- d_4) =\mbox{RHS}
\end{eqnarray*}
Note that the scheme is designed to satisfy server symmetry, so redundancy exists for Server 2 as well. Note also that the redundant symbols are created in the message symmetry step so that regardless of the value of $\theta$, the sign structure (alternating) is maintained and the symbol index structure is guaranteed to be symmetric.
So for all $\theta \in [1:4]$, we always have 3 redundant symbols from each server, and downloading 12 symbols per server suffices. The rate achieved is $L/D = 16/24 = 2/3 = C$.

\subsection{Proof of Optimality for Arbitrary $K, M$ and $N = 2$}\label{sec:proof_corr}
To prove optimality, we need to show that the scheme achieves a rate of $\left(1+1/2+\cdots+1/2^{K-1}\right)^{-1} = \frac{2^K}{2(2^{K} - 1)}$. For this, we will show that the user downloads only $\sum_{m=1}^M \left(\binom{M}{m} - \binom{M-K}{m}\right) = 2^M - 2^{M-K}$ symbols from each server. 
Note that the message size is $L = 2^M$, then the rate achieved is $\frac{2^M}{2(2^M - 2^{M-K})} =  \frac{2^K}{2(2^{K} - 1)}$, as desired.
Note that there are $\binom{M}{m}$ symbols queried in Block $m, m \in [1:M]$ from each server. However, it turns out that based on information available from the other sever, $\binom{M-K}{m}$ of these symbols are redundant. Thus, $\binom{M}{m} - \binom{M-K}{m}$ generic combinations of these $\binom{M}{m}$ symbols are sufficient.

Next we prove why this is the case in the following lemma.
\begin{lemma}\label{lemma:red}
For all $\theta \in [1:M]$, for each server, in Block $m \in [1:M-K]$, $\binom{M-K}{m}$ of the $\binom{M}{m}$ symbols are redundant, based on the information available from the other server.
\end{lemma}
{\it Proof:} Let us start with the case where $\theta = 1$. Consider the $m$-sum queries that do not involve the desired message $u_1$. There are $\binom{M-1}{m}$ such queries, divided into two groups:
\begin{enumerate}
\item  $\binom{M-1}{m} - \binom{M-K}{m}$ queries that involve at least one element in $\{u_2, \cdots, u_{K}\}$,
\item  $\binom{M-K}{m}$ queries that do not involve any element in $\{u_2, \cdots, u_{K}\}$.
\end{enumerate}

The key is that the symbols in Group 2 are redundant. Specifically, we show that they are functions of the symbols in Group 1 when $u_1$ is known.\footnote{This is guaranteed because the desired variable $u_1$ in Block $k$ is mixed with side information in Block $k-1$ available from the other server.}

{\it Example 1: We accompany the general proof with a concrete example to explain the idea. For this example, assume $K = 3$ datasets, $M = 6$ messages, and denote symbols $u_1, u_2, \cdots, u_6$ by distinct letters $a,b,\cdots, f$, respectively, for simplicity. Consider Block $m = 3$. The queries that do not involve the desired message $u_1$ are shown below.  For this example, we will see that the only symbol in Group 2 is a function of the 9 symbols in Group 1.
\begin{eqnarray*}
\begin{array}{|cc|c}\hline
\m{Group 1} & b_{j_5} - c_{j_2} + d_{j_1}  \\ 
& b_{j_6} - c_{j_3} + e_{j_1}  \\  
& b_{j_7} - c_{j_4} + f_{j_1}  \\
& b_{j_8} - d_{j_3} + e_{j_2}  \\
& b_{j_9} - d_{j_4} + f_{j_2}  \\
& b_{j_{10}} - e_{j_4} + f_{j_3}  \\  
& c_{j_8} - d_{j_6} + e_{j_5}  \\ 
& c_{j_9} - d_{j_7} + f_{j_5}  \\  
& c_{j_{10}} - e_{j_7} + f_{j_6}  \\ \hdashline
\m{Group 2} & d_{j_{10}} - e_{j_9} + f_{j_8}  \\
\hline
\end{array}~~
\end{eqnarray*}
}

To simplify the notation, define 
\begin{eqnarray}
q(u_{i_{[1:m]}}) &=& q(\{u_{i_1}, u_{i_2}, \cdots, u_{i_m}\}) \triangleq \sum_{l=1}^m (-1)^{l-1} u_{i_l}
\end{eqnarray}
where the message indices $i_1 < i_2 \cdots < i_m$, and the symbol indices are suppressed. 
Consider an arbitrary query in Group 2:
\begin{eqnarray*}
q_0 = q(u_{i_{[1:m]}}) 
\end{eqnarray*}
where $K < i_1 < i_2 \cdots < i_m$. We show that when $u_1$ is known, the query $q_0$ is a function of $\binom{m+K-1}{m} - 1$ queries in Group 1. These $\binom{m+K-1}{m} - 1$ queries contain an $m$-sum of every type in $\mathcal{I} \triangleq [2:K] \cup i_{[1:m]}$ (except $i_{[1:m]}$). 
\begin{eqnarray}
\mathcal{Q} \triangleq \Big\{ q(u_{j_{[1:m]}}): ~~ {j_{[1:m]}} \subset \mathcal{T} \Big\}
\end{eqnarray}
where the set of all possible $m$ distinct indices (types of $m$-sums) in $\mathcal{I}$ except ${i_{[1:m]}}$ is denoted as $\mathcal{T}$. Without loss of generality, we assume $j_1 < j_2 < \cdots < j_m$.
The indices of these queries are assigned by the index assignment process. 

From the linear dependence of the messages (\ref{linear}), we have
\begin{eqnarray}
u_{i_l} (*) &=& v_{i_{l}(1)} u_1(*) + \cdots + v_{i_{l}(K)} u_K(*), l \in [1:m]  \label{eq:u1}
\end{eqnarray}
Note that $u_1(*)$ are assumed known, so we may set $u_1(*)$ to zero. 
Now we show that $q_0$ is a linear function of the queries in $\mathcal{Q}$.
\begin{eqnarray}
q_0 = \sum_{{j_{[1:m]}} \in \mathcal{T}} h(u_{j_{[1:m]}}) q(u_{j_{[1:m]}}) \label{eq:q0}
\end{eqnarray}
where the linear combining coefficients $h(u_{j_{[1:m]}})$ are functions of ${\bf v}_{i_1}, \cdots, {\bf v}_{i_m}$. The elements of the matrix ${\bf V}^* \triangleq ({\bf v}_{i_1}^T ~{\bf v}_{i_2}^{T} ~\cdots ~ {\bf v}_{i_m}^{T})$ are shown below (the rows and columns are labelled by corresponding messages).
\[ {\bf V}^* =
   \left( \begin{array}{ccccc}
         {\bf v}_{i_1}^T & {\bf v}_{i_2}^{T} & \cdots & {\bf v}_{i_m}^{T} \\ 
          \end{array} \right) = 
    \begin{blockarray}{ccccc}
        & u_{i_1} & u_{i_2} & \cdots & u_{i_m} \\
      \begin{block}{c(cccc)}
        u_2 & v_{i_1(2)} & v_{i_2(2)} & \cdots & v_{i_m(2)} \\ 
        u_3 & v_{i_1(3)} & v_{i_2(3)} & \cdots & v_{i_m(3)} \\
        \vdots & \vdots & \vdots & \ddots & \vdots \\
        u_K & v_{i_1(K)} & v_{i_2(K)} & \cdots & v_{i_m(K)} \\
      \end{block}
    \end{blockarray}
 \]
In particular, $h(u_{j_{[1:m]}})$ are specified as follows. Suppose $|j_{[1:m]} \cap [2:K]| = t$, where $t \in [1:m]$ and denote these $t$ elements as $\bar{j}_{[1:t]} \triangleq j_{[1:m]} \cap [2:K]$.
Then $|j_{[1:m]} \cap i_{[1:m]}| = m - t$ and denote these $m-t$ elements as $\bar{i}_{[1:m-t]} \triangleq j_{[1:m]} \cap i_{[1:m]}$. We further define $\widetilde{i}_{[1:t]} \triangleq i_{[1:m]} / \bar{i}_{[1:m-t]}$, where $\widetilde{i}_{1} < \cdots < \widetilde{i}_{t}$. 
We are now ready to give $h(u_{j_{[1:m]}})$. $h(u_{j_{[1:m]}})$ is equal to the determinant of the {\color{black} $t\times t$ square matrix obtained as the sub-matrix of ${\bf V}^*$ }where the rows correspond to messages $u_{\bar{j}_{[1:t]}}$ and the columns correspond to messages $u_{\widetilde{i}_{[1:t]}}$.
\begin{eqnarray}
h(u_{j_{[1:m]}}) = (-1)^{\sum_{r=1}^t \Omega(\widetilde{i}_{r})+t(t-1)/2+1}
\left\rvert
\begin{array}{ccccc}
        v_{\widetilde{i}_1(\bar{j}_1)} & v_{\widetilde{i}_2(\bar{j}_1)} & \cdots & v_{\widetilde{i}_t(\bar{j}_1)} \\ 
        v_{\widetilde{i}_1(\bar{j}_2)} & v_{\widetilde{i}_2(\bar{j}_2)} & \cdots & v_{\widetilde{i}_t(\bar{j}_2)} \\
        \vdots & \vdots & \ddots & \vdots \\
        v_{\widetilde{i}_1(\bar{j}_t)} & v_{\widetilde{i}_2(\bar{j}_t)} & \cdots & v_{\widetilde{i}_t(\bar{j}_t)} \\
\end{array}
\right\rvert
\label{eq:h1}
\end{eqnarray}
where $\Omega(\widetilde{i}_{r})$ is defined as the position of $\widetilde{i}_r$ in the lexicographic ordering of the elements of $i_{[1:m]}$. {\color{black}For example, suppose $i_{[1:k]} = \{4, 6 , 7, 9\}$. Then if $\widetilde{i}_r = 6$, then $\Omega(\widetilde{i}_{r}) = 2$.  Similarly,  if $\widetilde{i}_r = 9$, then $\Omega(\widetilde{i}_{r}) = 4$.}

Let us verify that (\ref{eq:q0}) holds. In (\ref{eq:q0}), $\binom{m+K-1}{m-1}$ distinct {\color{black} symbol} indices appear, {\color{black} and} each {\color{black} of those symbol indices} is assigned to $K$ {\color{black} message} variables. Pick any $m-1$ messages from the $m + K -1$ messages $u_{\mathcal{I}}$, say $u_{\alpha_{[1:m-1]}}$, where $\alpha_1 < \cdots< \alpha_{t} \leq K < \alpha_{t+1} < \cdots < \alpha_{m-1}, t \in [0:K-1]$. 
The same index (denoted by $\#$) is assigned to the variables 
\begin{eqnarray}
u_{\mathcal{I}}/u_{\alpha_{[1:m-1]}} \triangleq u_{\beta_{[1:K]}} \label{eq:ui}
\end{eqnarray}
where $\beta_1 < \cdots< \beta_{K-1-t} \leq K < \beta_{K-t} < \cdots < \beta_{K}$. From (\ref{eq:ui}), we have
\begin{eqnarray}
&& \alpha_{[1:t]} \cup \beta_{[1:K-1-t]}= [2:K] \\
&& \alpha_{[t+1:m-1]} \cup \beta_{[K-t:K]} = i_{[1:m]}
\end{eqnarray}
The $K$ variables $u_{\beta_{[1:K]}}(\#)$ 
appear in the following $K$ queries.
\begin{eqnarray}
q_l \triangleq q(u_{\alpha_{[1:m-1]} \cup \beta_l}), l \in [1:K].
\end{eqnarray}

We show that for any $m-1$ distinct indices $\alpha_{[1:m-1]}$ in $\mathcal{I}$, (\ref{eq:q0}) holds for the $K$ variables $u_{\beta_{[1:K]}}(\#)$.
Using (\ref{eq:u1}), we write $u_{\beta_{[1:K]}}(\#)$ as linear combinations of $u_{[2:K]}(\#)$. Next we prove that (\ref{eq:q0}) holds for $u_\eta(\#), \forall \eta \in [2:K]$. 
Define
\begin{eqnarray}
{\bf V} = [V_{i,j}]_{(t+1) \times (t+1)} \triangleq \left( \begin{array}{cccccc}
        v_{\beta_{K-t}(\eta)} & v_{\beta_{K-t+1}(\eta)} & \cdots & v_{\beta_{K}(\eta)} \\ 
        v_{\beta_{K-t}(\alpha_1)} & v_{\beta_{K-t+1}(\alpha_1)} & \cdots & v_{\beta_{K}(\alpha_1)} \\ 
        v_{\beta_{K-t}(\alpha_2)} & v_{\beta_{K-t+1}(\alpha_2)} & \cdots & v_{\beta_{K}(\alpha_2)} \\ 
        \vdots & \vdots & \ddots & \vdots \\
        v_{\beta_{K-t}(\alpha_t)} & v_{\beta_{K-t+1}(\alpha_t)} & \cdots & v_{\beta_{K}(\alpha_t)} \\ 
\end{array} \right)
\end{eqnarray}
and the minor of ${\bf V}$ (the determinant of the submatrix formed by deleting the $i$-th row and $j$-column) is denoted by $M_{i,j}$.
Note that $\alpha_{[t+1 : m-1]} \cup \beta_{[K-t : K]} = i_{[1:K]}$, so
\begin{eqnarray}
\{\Omega(\alpha_{t+1}) \cup \cdots \cup \Omega(\alpha_{m-1}) \cup \Omega(\beta_{K-t}) \cup \cdots \Omega(\beta_K) \} = \{\Omega(i_{1}) \cup \cdots \cup \Omega(i_K)\} = 1:K
\end{eqnarray}
and
\begin{eqnarray}
\Delta_{u_{\beta_r}}\left(q_r \right) = t + \Omega(\beta_{\gamma}) - \left(r - (K-t)\right), \forall r \in [K-t:K] \label{eq:lo}
\end{eqnarray}

We now consider  two cases for $\eta$.
\begin{enumerate}
\item[Case $1$:] $\eta \in \alpha_{[1:t]}$. In this case, $u_\eta(\#)$ variables come from $u_{\beta_{[K-t:K]}}(\#)$. (\ref{eq:q0}) boils down to 
\begin{eqnarray}
&&\left(\sum_{r = K-t}^K  h(u_{\alpha_{[1:m-1]} \cup \beta_{r}}) \times (-1)^{\Delta_{u_{\beta_r}}\left(q_r \right)+1} v_{\beta_{r}(\eta)} \right) \times u_\eta(\#) = 0\\
&\Longleftarrow& 
        v_{\beta_{K-t}(\eta)} (-1)^{\Delta_{u_{\beta_{K-t}}}\left(q_{K-t} \right)+1} (-1)^{\sum_{s = K-t}^K \Omega(\beta_s) - \Omega(\beta_{K-t})+t(t-1)/2+1} M_{1,1} \notag \\
        &&+ v_{\beta_{K-t+1}(\eta)} (-1)^{\Delta_{u_{\beta_{K-t+1}}}\left(q_{K-t+1} \right)+1} (-1)^{\sum_{s = K-t}^K \Omega(\beta_s) - \Omega(\beta_{K-t+1}) +t(t-1)/2+1} M_{1,2}+ \cdots \notag \\
        &&+ v_{\beta_{K}(\eta)} (-1)^{\Delta_{u_{\beta_{K}}}\left(q_{K} \right)+1} (-1)^{\sum_{s = K-t}^K \Omega(\beta_s) - \Omega(\beta_K)+t(t-1)/2+1} M_{1, t+1} = 0 \label{eq:as1} \\
&\Longleftarrow& 
        v_{\beta_{K-t}(\eta)} M_{1,1} - v_{\beta_{K-t+1}(\eta)} M_{1,2} \cdots + (-1)^{t+2} v_{\beta_{K}(\eta)} M_{1, t+1} = 0 \label{eq:as2}
        \\
&\Longleftarrow& 
        V_{1,1} M_{1,1} - V_{1,2} M_{1,2} \cdots + (-1)^{t+2} V_{1,t+1} M_{1, t+1} 
        = |{\bf V}| = 0 \label{eq:vu}
\end{eqnarray}
where (\ref{eq:as2}) follows from the observation that consecutive terms in (\ref{eq:as1}) have alternating signs, proved as follows. For any $r \in [K-t:K-1]$, 
\begin{eqnarray}
&& (-1)^{\Delta_{u_{\beta_r}}\left(q_{r} \right)+1} (-1)^{\sum_{s = K-t}^K \Omega(\beta_s) - \Omega(\beta_{r})+t(t-1)/2+1} \notag \\
&\overset{(\ref{eq:lo})}{=}& (-1)^{t + \Omega_{\beta_r} - \left(r - (K-t)\right)+1} (-1)^{\sum_{s = K-t}^K \Omega(\beta_s) - \Omega(\beta_{r})+t(t-1)/2+1} \notag \\
&=& (-1)^{t - \left(r - (K-t)\right) +1} (-1)^{\sum_{s = K-t}^K \Omega(\beta_s)+t(t-1)/2+1} \notag \\
&=& (-1)\times (-1)^{t + \Omega(\beta_{r+1}) - \left(r+1 - (K-t)\right)+1} (-1)^{\sum_{s = K-t}^{K} \Omega({\beta_s}) - \Omega(\beta_{r+1}) +t(t-1)/2+1} \notag \\
&\overset{(\ref{eq:lo})}{=}& (-1) \times (-1)^{\Delta_{u_{\beta_{r+1}}}\left(q_{r+1} \right)+1} (-1)^{\sum_{s = K-t}^K \Omega(\beta_s) - \Omega(\beta_{r+1})+t(t-1)/2+1}
\end{eqnarray}
(\ref{eq:vu}) is due to the fact that $\eta \in \alpha_{[1:t]}$, so ${\bf V}$ has two identical rows and its determinant is 0.
\item [Case $2$:] $\eta \in \beta_{[1:K-1-t]}$. In this case, $u_\eta(\#)$ variables come from $u_{\beta_{[K-t:K]} \cup \eta}(\#)$. If $\alpha_{[1:m-1]} \cap [2:K] \neq \emptyset$, (\ref{eq:q0}) boils down to
\begin{eqnarray}
&&\left(h(u_{\eta\cup \alpha_{[1:m-1]}}) (-1)^{\Delta_{u_\eta}(q(u_{\alpha_{[1:m-1]} \cup \eta}))+1} + \sum_{r = K-t}^K h(u_{\alpha_{[1:m-1]} \cup \beta_{r}}) \times (-1)^{\Delta_{u_{\beta_r}}\left(q_r \right)+1} v_{\beta_{r}(\eta)} \right) \notag\\
&&~\times u_\eta(\#) = 0 \label{eq:vv1} \\
&\Longleftarrow& |{\bf V}| - |{\bf V}| = 0 \label{eq:vv}
\end{eqnarray}
where the second term of (\ref{eq:vv}) follows from (\ref{eq:vu}) and the `$-$' sign in (\ref{eq:vv}) is due to the fact that in (\ref{eq:vv1}), the sign of the first term is different from the sign of the second term, proved as follows.
\begin{eqnarray}
&& (-1)^{\Delta_{u_\eta}(q(u_{\alpha_{[1:m-1]} \cup \eta}))+1} (-1)^{\Delta_{u_\eta}(q(u_{\alpha_{[1:m-1]} \cup \eta}))+1} (-1)^{\sum_{s=K-t}^K \Omega(\beta_s) +t(t+1)/2+ 1}\notag\\
&=& (-1) \times (-1)^{t + \Omega(\beta_{K-t}) + 1} (-1)^{\sum_{s = K-t+1}^{K} \Omega(\beta_s)+t(t-1)/2+1} \notag \\
&\overset{(\ref{eq:lo})}{=}& (-1) \times (-1)^{\Delta_{u_{\beta_{K-t}}}\left(q_{K-t} \right)+1} (-1)^{\sum_{s = K-t+1}^K \Omega(\beta_s) +t(t-1)/2+1}
\end{eqnarray}
\end{enumerate}
Note that in the first line, the first $(-1)^{\Delta_{u_\eta}(q(u_{\alpha_{[1:m-1]} \cup \eta}))+1}$ term is to account for the different ordering of the vectors in ${\bf V}$ that appear in defining $h(u_{\eta\cup \alpha_{[1:m-1]}})$.

Otherwise, if $\alpha_{[1:m-1]} \cap [2:K] = \emptyset$, i.e., $\alpha_{[1:m-1]} \subset i_{[1:m]}$, we have $t = 0$ and (\ref{eq:q0}) boils down to
\begin{eqnarray}
&& (-1)^{\Delta_{u_{\beta_K}}\left(q_K \right)+1} v_{\beta_{K(\eta)}} = h(u_{\eta\cup \alpha_{[1:m-1]}}) (-1)^{\Delta_{u_\eta}(q(u_{\alpha_{[1:m-1]} \cup \eta}))+1}  \\
&\Longleftarrow& ~~~ (-1)^{\omega(\beta_K)+1} v_{\beta_{K(\eta)}} = h(u_{\eta\cup \alpha_{[1:m-1]}})  \label{eq:l2} 
\end{eqnarray}
where (\ref{eq:l2}) follows from $\Delta_{u_{\beta_K}}\left(q_K \right) = \omega(\beta_K)$ as in $q_K$, the messages are $u_{\beta_K} \cup u_{\alpha_{[1:m-1]}} = u_{i_{[1:m]}}$, and $\Delta_{u_\eta}(q(u_{\alpha_{[1:m-1]} \cup \eta})) = 1$ as $\eta \leq K < \alpha_1$. Note that (\ref{eq:l2}) is the definition of $h(u_{\eta\cup \alpha_{[1:m-1]}})$ (see (\ref{eq:h1})).
Therefore the proof is complete.

{\it Example 1 (Continued): Consider the query in Group 2, $d_{j_{10}} - e_{j_9} + f_{j_8}$. 
We show that it is a function of the $9$ queries in Group 1, when the desired variables ($a_*$) are set to zero.
\begin{eqnarray*}
&& d_{j_{10}} - e_{j_9} + f_{j_8} \\
&=& -\left\rvert
\begin{array}{cc}
v_{5(2)} & v_{6(2)} \\
v_{5(3)} & v_{6(3)} \\
\end{array}
\right\rvert
\left( b_{j_5} - c_{j_2} + d_{j_1} \right)  + \left\rvert
\begin{array}{cc}
v_{4(2)} & v_{6(2)} \\
v_{4(3)} & v_{6(3)} \\
\end{array}
\right\rvert 
\left( b_{j_6} - c_{j_3} + e_{j_1} \right)  - \left\rvert
\begin{array}{cc}
v_{4(2)} & v_{5(2)} \\
v_{4(3)} & v_{5(3)} \\
\end{array}
\right\rvert
 \left( b_{j_7} - c_{j_4} + f_{j_1} \right) \\
&& +~ v_{6(2)} \left( b_{j_8} - d_{j_3} + e_{j_2} \right)  - v_{5(2)} \left( b_{j_9} - d_{j_4} + f_{j_2} \right) + v_{4(2)} \left( b_{j_{10}} - e_{j_4} + f_{j_3} \right)  \\  
&& +~ v_{6(3)} \left( c_{j_8} - d_{j_6} + e_{j_5} \right) - v_{5(3)} \left( c_{j_9} - d_{j_7} + f_{j_5} \right) + v_{4(3)} \left( c_{j_{10}} - e_{j_7} + f_{j_6} \right)
\end{eqnarray*}
}

{\color{black}
{\it Example 2: Let us include another example, where $K = 4$, $M = 8$. Consider Block $m = 3$ and the desired message index $\theta = 1$. The queries that do not involve $u_1$ are divided into Group 1 (where $u_2, u_3$ or $u_4$ appears) and Group 2 (where none of $u_2, u_3, u_4$ appears). Consider a query in Group 2, $q_0 = q(u_{5,6,8})$, i.e., $i_1 = 5, i_2 = 6, i_3 = 8$. When $u_1$ is {\color{black}known}, $q_0$ is a function of the following $\binom{3+4-1}{3} - 1 = 19$ queries. Here $\mathcal{I} = \{2,3,4,5,6,8\}$.
\begin{eqnarray}
\mathcal{Q} &=& \Big\{ q(u_{2,3,4}), q(u_{2,3,5}), q(u_{2,3,6}), q(u_{2,3,8}), q(u_{2,4,5}), q(u_{2,4,6}), q_(u_{2,4,8}), q_(u_{2,5,6}), q_(u_{2,5,8}), q(u_{2,6,8}), \notag\\
&&~~ q(u_{3,4,5}), q(u_{3,4,6}), q(u_{3,4,8}), q(u_{3,5,6}), q(u_{3,5,8}), q(u_{3,6,8}), q(u_{4,5,6}), q(u_{4,5,8}), q(u_{4,6,8})  \Big\}
\end{eqnarray}
The linear combining coefficients in (\ref{eq:q0}) are designed following (\ref{eq:h1}). Let us verify (\ref{eq:q0}) for the symbols with a particular index value, $\#$. To this end, let us pick {\color{black} the $m-1=2$ message indices} $\alpha_1 = 3, \alpha_2 = 4$ (note that $\{3,4\} \subset \mathcal{I}$). As $\alpha_2 = 4\leq K =4$, we have $t = 2$. The variables with index $\#$ are from $u_2, u_5, u_6, u_8$ (from the difference set of $\mathcal{I}$ and $\{\alpha_1, \alpha_2\}$), so that we have $\beta_1 = 2, \beta_2 = 5, \beta_3 = 6, \beta_4 = 8$. These 4 variables appear in queries 
\begin{eqnarray}
q_1 = q(u_{2,3,4}), q_2 = q(u_{3,4,5}), q_3 = q(u_{3,4,6}), q_4 = q(u_{3,4,8})
\end{eqnarray}
We can write $u_5(\#), u_6(\#), u_8(\#)$ as a linear combination of $u_2(\#), u_3(\#), u_4(\#)$ {\color{black}after $u_1(\#)$ is eliminated, or equivalently, set to zero}.
Next we show that (\ref{eq:q0}) holds for $u_3(\#)$. In this case, $\eta = 3$ and $\eta \subset \{\alpha_1, \alpha_2\} = \{3,4\}$, {\color{black}so we are in  Case $1$}. We want to show the following.
\begin{eqnarray}
&&\Big( h(u_{3,4,5}) \times (-1)^{\Delta_{u_5}(q(u_{3,4,5})) + 1} v_{5(3)} + h(u_{3,4,6}) \times (-1)^{\Delta_{u_6}(q(u_{3,4,6})) + 1} v_{6(3)} \notag \\
&&+~ h(u_{3,4,8}) \times (-1)^{\Delta_{u_8}(q(u_{3,4,8})) + 1} v_{8(3)} \Big) u_3(\#)= 0 \\
&\Longleftrightarrow& h(u_{3,4,5}) v_{5(3)} + h(u_{3,4,6})v_{6(3)} + h(u_{3,4,8}) v_{8(3)} = 0 \label{eq:hd1}
\end{eqnarray} 
Note that $\Delta_{u_5}(q(u_{3,4,5}))$ is related to $\Omega{(5)}$.
We now find $h(u_{3,4,5})$. Referring to (\ref{eq:h1}), we have
\begin{eqnarray}
&&j_1 = 3, j_2 = 4, j_3 = 5, \bar{j}_1 = 3, \bar{j}_2 = 4 \\
&&\bar{i} = 5, \widetilde{i}_1 = 6, \widetilde{i}_2 = 8, \Omega(6) = 2, \Omega(8) = 3\\
&& h(u_{3,4,5}) = (-1)^{2+3 + 2\times1/2 + 1} \left\rvert
\begin{array}{cc}
v_{6(3)} & v_{8(3)} \\
v_{6(4)} & v_{8(4)} \\
\end{array}
\right\rvert   = - \left\rvert
\begin{array}{cc}
v_{6(3)} & v_{8(3)} \\
v_{6(4)} & v_{8(4)} \\
\end{array}
\right\rvert 
\end{eqnarray}
Similarly, 
\begin{eqnarray}
h(u_{3,4,6}) =  \left\rvert
\begin{array}{cc}
v_{5(3)} & v_{8(3)} \\
v_{5(4)} & v_{8(4)} \\
\end{array}
\right\rvert,
h(u_{3,4,6}) = - \left\rvert
\begin{array}{cc}
v_{5(3)} & v_{6(3)} \\
v_{5(4)} & v_{6(4)} \\
\end{array}
\right\rvert
\end{eqnarray}
Therefore (\ref{eq:hd1}) is equivalent to
\begin{eqnarray}
- \left\rvert
\begin{array}{cc}
v_{6(3)} & v_{8(3)} \\
v_{6(4)} & v_{8(4)} \\
\end{array}
\right\rvert v_{5(3)}
+
 \left\rvert
\begin{array}{cc}
v_{5(3)} & v_{8(3)} \\
v_{5(4)} & v_{8(4)} \\
\end{array}
\right\rvert v_{6(3)}
- \left\rvert
\begin{array}{cc}
v_{5(3)} & v_{6(3)} \\
v_{5(4)} & v_{6(4)} \\
\end{array}
\right\rvert v_{8(3)} = 
- \left\rvert
\begin{array}{ccc}
v_{5(3)} & v_{6(3)} & v_{8(3)}\\
v_{5(3)} & v_{6(3)} & v_{8(3)}\\
v_{5(4)} & v_{6(4)} & v_{8(4)} \\
\end{array}
\right\rvert = 0
\end{eqnarray}
and thus (\ref{eq:hd1}) holds.
For the other case (Case $2$), we show that (\ref{eq:q0}) holds for $u_2(\#)$, i.e., $\eta = 2$ and $\eta = \beta_1 = 2$. In this case, we want to show
\begin{eqnarray}
&&\Big( h(u_{2,3,4}) \times (-1)^{\Delta_{u_2}(q(u_{2,3,4})) + 1} + h(u_{3,4,5}) \times (-1)^{\Delta_{u_5}(q(u_{3,4,5})) + 1} v_{5(2)} \notag\\
&&+~ h(u_{3,4,6}) \times (-1)^{\Delta_{u_6}(q(u_{3,4,6})) + 1} v_{6(2)} + h(u_{3,4,8}) \times (-1)^{\Delta_{u_8}(q(u_{3,4,8})) + 1} v_{8(2)} \Big) \notag\\
&&\times u_2(\#) = 0 \\
&\Longleftrightarrow& h(u_{2,3,4}) + h(u_{3,4,5}) v_{5(2)} + h(u_{3,4,6})v_{6(2)} + h(u_{3,4,8}) v_{8(2)} = 0 \label{eq:hd2}
\end{eqnarray} 
Following the definition of $h(u_{2,3,4})$ (refer to (\ref{eq:h1})), we find that
\begin{eqnarray}
h(u_{2,3,4}) = \left\rvert
\begin{array}{ccc}
v_{5(2)} & v_{6(2)} & v_{8(2)}\\
v_{5(3)} & v_{6(3)} & v_{8(3)}\\
v_{5(4)} & v_{6(4)} & v_{8(4)} \\
\end{array}
\right\rvert 
\end{eqnarray}
Then (\ref{eq:hd2}) is equivalent to
\begin{eqnarray}
 \left\rvert
\begin{array}{ccc}
v_{5(2)} & v_{6(2)} & v_{8(2)}\\
v_{5(3)} & v_{6(3)} & v_{8(3)}\\
v_{5(4)} & v_{6(4)} & v_{8(4)} \\
\end{array}
\right\rvert  -  \left\rvert
\begin{array}{ccc}
v_{5(2)} & v_{6(2)} & v_{8(2)}\\
v_{5(3)} & v_{6(3)} & v_{8(3)}\\
v_{5(4)} & v_{6(4)} & v_{8(4)} \\
\end{array}
\right\rvert  = 0
\end{eqnarray}
and thus (\ref{eq:hd2}) holds.
Let us consider another index ($\#'$) where $\alpha_1 = 5, \alpha_2 = 6$, i.e., $\alpha_1 > K =4$ and $t = 0$. The index $\#'$ is assigned to variables from $u_2, u_3, u_4, u_7$ ($\beta_1 = 2, \beta_2 = 3, \beta_3 = 4, \beta_4 = 7$) in queries
\begin{eqnarray}
q_1 = q(u_{2,5,6}), q_2 = q(u_{3,5,6}), q_3 = q(u_{4,5,6}), q_4 = q(u_{5,6,7})
\end{eqnarray}
After writing every variable in terms of $u_2, u_3, u_4$ {\color{black}($u_1$ terms are set to zero because they are known and can be removed)}, we show that (\ref{eq:q0}) holds for $u_2(\#'), u_3(\#'), u_4(\#')$. Note that no matter which variable we pick, say $u_4(\#')$, i.e., $\eta = 4$, $\eta \in \{2,3,4\} = \{\beta_1, \beta_2, \beta_3\}$. Further $\{\alpha_1, \alpha_2\} \cap \{2,3,4\} = \emptyset$. In this case, we want to show
\begin{eqnarray}
&&(-1)^{\Delta_{u_{7}}\left(q_{5,6,7} \right)+1} v_{7(4)} u_4(\#') = h(u_{4,5,6}) (-1)^{\Delta_{u_4}(q(u_{4,5,6})+1} u_4(\#')\\
&\Longleftrightarrow& v_{7(4)} = h(u_{4,5,6})
\end{eqnarray}
which matches the definition of $h(u_{4,5,6})$ (see (\ref{eq:h1})) thus holds.
}
}

The proof for arbitrary $\theta \neq 1$ follows similarly. 
Since the first $K$ of the $M$ linear combinations are linearly independent (in fact, they are the $K$ independent datasets), there exist $K-1$ messages from $u_{[1:K]}$ (denoted as $u_{r_{[2:K]}}, r_{[2:K]} \subset [1:K]$) such that $u_\theta \cup u_{r_{[2:K]}}$ {\color{black} are independent}. 
Similarly, consider the $m$-sum queries that do not involve the desired message $u_\theta$, which are further divided into two groups, depending on whether at least one element from $u_{r_{[2:K]}}$ is involved (Group 1) or not (Group 2). 
We show that any query $q_0 = q(u_{i_{[1:m]}}), i_{[1:m]} \cap (\theta \cup r_{[2:K]}) = \emptyset$ in Group 2 is a function of the queries in Group 1. $q_0$ exists as $m \leq M-K$.
The symbol indices in $q_0$ are assigned by the index assignment process.
By a change of basis, we express each variable as a linear combination of $u_\theta \cup u_{r_{[2:K]}}$. 
Then we show that $q_0$ is a linear combination of the queries $q(u_{j_{[1:m]}})$, where ${j_{[1:m]}} \in \mathcal{T}'$, and 
$\mathcal{T}'$ is the set of all possible $m$ distinct indices in $r_{[2:K]} \cup i_{[1:m]}$ except $i_{[1:m]}$.
The rest of the proof,  where we design the linear combining coefficients and show the linear combination holds, is identical to the case of $\theta = 1$ (by an invertible mapping from $r_{[2:K]}$ to $[2:K]$, and between $i_{[1:m]}$ of the two cases). 

{\color{black}
{\it Example 3: We give an example where $\theta \neq 1$. Assume $K = 3$ datasets, $M = 6$ messages, $\theta = 5$, and denote symbols $u_1, u_2, \cdots, u_6$ by distinct letters $a,b,\cdots, f$, respectively. Consider Block $m = 2$. There exists two messages in $a, b, c$ (assume without loss of generality, $a, b$) such that $a, b, e$ are independent. The queries that do not involve the desired message $e$ are shown below.  The queries are divided into Group 1 (where $a$ or $b$ appears) and Group 2 (where none of $a, b$ appears).
\begin{eqnarray*}
\begin{array}{|cc|c}\hline
\m{Group 1} & a_{j_2} - b_{j_1}\\
&a_{j_3} - c_{j_1} \\
&a_{j_4} - d_{j_1} \\
&a_{j_5} - f_{j_1} \\
&b_{j_3} - c_{j_2} \\
&b_{j_4} - d_{j_2} \\
&b_{j_5} - f_{j_2}  \\ \hdashline
\m{Group 2} & c_{j_4} - d_{j_3} \\
&c_{j_5} - f_{j_3} \\
&d_{j_5} - f_{j_4} \\
\hline
\end{array}~~
\end{eqnarray*}
We express $c, d, f$ as a linear combination of $a, b, e$ (note that $a, b, e$ are linearly independent). Assume
\begin{eqnarray}
c &=& v_{c(a)}a + v_{c(b)} b + v_{c(e)} e \\
d &=& v_{d(a)}a + v_{d(b)} b + v_{d(e)} e \\
f &=& v_{f(a)}a + v_{f(b)} b + v_{f(e)} e 
\end{eqnarray}
The queries in Group 2 are functions of the queries in Group 1. For example, consider $c_{j_5} - f_{j_3}$. When $e_*$ are set to zero, we have
\begin{eqnarray}
c_{j_5} - f_{j_3} &=& -\left\rvert
\begin{array}{cc}
v_{c(a)} & v_{f(a)} \\
v_{c(b)} & v_{f(b)} \\
\end{array}
\right\rvert  \left( a_{j_2} - b_{j_1} \right) - v_{f(a)} \left( a_{j_3} - c_{j_1} \right) + v_{c(a)} \left( a_{j_5} - f_{j_1} \right) \notag\\
&&-~ v_{f(b)} \left( b_{j_3} - c_{j_2} \right) + v_{c(b)}\left( b_{j_5} - f_{j_2}\right) 
\end{eqnarray}
where the linear combining coefficients are determined by the following matrix.
\[ 
    \begin{blockarray}{ccc}
        & c & f \\
      \begin{block}{c(cc)}
        a& v_{c(a)} & v_{f(a)}  \\ 
        b & v_{c(b)} & v_{f(b)} \\
      \end{block}
    \end{blockarray}
 \]
 For example, for $a_{j_3} - c_{j_1}$, from (\ref{eq:h1}), the linear coefficient is $(-1)^{2+0+1} v_{f(a)} = - v_{f(a)}$.
}
}

\subsection{Proof of Optimality for Arbitrary $K, M, N$}\label{sec:proof_corr_general}
The proof of optimality when $N > 2$ follows from that when $N=2$. 
The query structure of any query vertex at level $m$ for arbitrary $N$ is identical to the structure of a query vertex at level $m$ for the  $N=2$ setting. From the observations listed in Section \ref{kmn}, recall that for any $N > 2$, the queries from each server in block $m$ are made up of $(N-1)^{m-1}$ query vertices.
Also let us recall from Lemma \ref{lemma:red} that when $N=2$, for each server there are $\binom{M-K}{m}$ redundant symbols within each level $m$ query vertex, $m\in[1:M-K]$. Therefore, when $N>2$, there are $(N-1)^{m-1} \binom{M-K}{m}$ redundant symbols in block $m$, and it suffices to download only $N\left( \sum_{m=1}^M  (N-1)^{m-1} \left( \binom{M}{m} - \binom{M-K}{m} \right) \right)$ symbols in total from all $N$ servers. The rate achieved is
\begin{eqnarray}
R &=& \frac{N^M}{N\left( \sum_{m=1}^M  (N-1)^{m-1} \left( \binom{M}{m} - \binom{M-K}{m} \right) \right)} \\
&=& \frac{N^M}{N \times \frac{1}{N-1} \left(N^M - N^{M-K}\right)} = \frac{N-1}{N} \frac{N^K}{N^K-1}\\
&=& \left( 1 + \frac{1}{N} + \cdots + \frac{1}{N^{K-1}} \right)^{-1}
\end{eqnarray}
which matches the capacity of private computation. The optimality proof is therefore complete.

\section{Proof of Privacy of $\PC$}
\subsection{Proof of Privacy for Example $A$ }\label{sec:privacy_toy}
To see why this scheme is private, we show that the queries are identically distributed, regardless of the value of $\theta$. To this end, we show that the query for $\theta = 2, 3, 4$ has a one-to-one mapping to the query for $\theta = 1$, respectively, through a choice of permutation $\pi$ and signs $\sigma_i$ which is made privately and uniformly by the user.

For example, for Server 1 and Server 2, the query for $\theta = 2$ can be converted into the query for $\theta = 1$ by the following mapping:
\begin{eqnarray*}
&\mbox{\footnotesize Server 1:}&(3,2, 7, 9, 10, 8, 15, 14, -\sigma_6, -\sigma_{12}, -\sigma_{13})\\
&\stackrel{}{\longrightarrow}& (2, 3, 9, 7, 8, 10,14, 15, \sigma_6, \sigma_{12}, \sigma_{13})\\
&\mbox{\footnotesize Server 2:}&(6,1,12, 4,13, 5,16, 11, -\sigma_3, -\sigma_{9}, -\sigma_{10})\\
&\stackrel{}{\longrightarrow}&(1,6, 4, 12,5,13,11, 16,\sigma_3, \sigma_9, \sigma_{10})
\end{eqnarray*}
However, these mappings are privately generated by the user and both alternatives are equally likely regardless of desired message. Hence, these queries are indistinguishable.

We can similarly verify that the other remaining queries for $\theta=3, 4$, are indistinguishable as well.
For Server 1 and Server 2, the query for $\theta=3$ can be converted into the query for $\theta=1$ by the following mapping: 
\begin{eqnarray*}
&\mbox{\footnotesize Server 1:}&(3,4,2,7,6,9,10,11,8, -\sigma_8,14,13,15, -\sigma_{12})\\
&\stackrel{}{\longrightarrow}& (2,3,4,9,7,6,8,10,11,\sigma_{11},15,14,13, \sigma_{12})\\
&\mbox{\footnotesize Server 2:}&(7,6,1,4,3,12,14,13,5,-\sigma_5,11,10,16, -\sigma_{9})\\
&\stackrel{}{\longrightarrow}&(6,1,7,12,4,3,13,5,14,\sigma_{14},16,11,10, \sigma_{9})
\end{eqnarray*}
The last case is when $\theta = 4$. The mapping from that to $\theta = 1$ is as follows.
\begin{eqnarray*}
&\mbox{\footnotesize Server 1:}&(3,4,5,2,6,7,8,9,10,11,14,13,12,15)\\
&\stackrel{}{\longrightarrow}& (2,3,4,5,8,10,11,6,7,9,15,14,13,12)\\
&\mbox{\footnotesize Server 2:}&(6,7,8,1,3,4,5,12,13,14,11,10,9,16)\\
&\stackrel{}{\longrightarrow}&(1,6,7,8,5,13,14,3,4,12,16,11,10,9)
\end{eqnarray*}
Again, since these mappings are privately generated by the user and both alternatives are equally likely regardless of desired message, these queries are indistinguishable. Thus all queries are indistinguishable and the scheme is private.

\subsection{Proof of Privacy for Arbitrary $K, M, N$}\label{sec:privacy_general}
We prove that $\PC$ is private. We know that $\PIRB$ is private and $\PC$ is obtained from $\PIRB$ by the sign assignment. Therefore it suffices to show that the sign assignment does not destroy privacy, i.e., $Q(n, \text{`}\theta\text{'})$ still has a one-to-one mapping to $Q(n, \text{`}1\text{'})$ by a choice of permutation $\pi$ and signs $\sigma_i$ which is made by the user privately and uniformly. 

The one-to-one mapping is quite simple. Note that each query in $Q(n, \text{`}1\text{'})$ has alternating signs. Consider $Q(n, \text{`}\theta\text{'})$. {\color{black}We only need to consider the non-desired symbols in queries introduced by {\bf Exploit-SI} (so $u_\theta$ is involved).
The reason is that the signs of the desired symbols introduced by {\bf Exploit-SI} and the other queries introduced by {\bf M-Sym} are the same as the signs of the queries in $Q(n, \text{`}1\text{'})$.}\footnote{\color{black}Note that the indices of the non-desired symbols introduced by {\bf Exploit-SI} do not appear in the queries introduced by {\bf M-Sym}. The reason is seen as follows. Consider a symbol $u_i, i \neq \theta$ that appears in a query introduced by {\bf Exploit-SI} (denote the query by $q$, so $u_\theta$ appears in $q$) and suppose the index of $u_i$ is $j$ (i.e., we have $u_i(j)$). Now from index assignment, symbols with index $j$ all appear in terms that contain $u_\theta$ (thus these terms are all generated by {\bf Exploit-SI}).} 
These queries all satisfy that $\Delta_{W_\theta} > 0$.
Now to map $Q(n, \text{`}\theta\text{'})$ to $Q(n, \text{`}1\text{'})$, for each block, we flip the signs (i.e., replace $\sigma_i$ with $-\sigma_i$) of variables to the right of $u_\theta$ in queries from sub-blocks $S$ if $S$ is odd, and the signs of variables to the left of $u_\theta$ in queries from sub-blocks $S$ if $S$ is even.

{\it Example 4: We accompany the general proof with a concrete example to explain the idea. Consider $M = 6$ (messages), block $m = 4$, desired message index $\theta = 4$. For simplicity, we denote $u_1, u_2, \cdots, u_6$ by $a, b, \cdots, f$.
In Block $B = m = 4$, we have $\binom{6-1}{4-1} = 10$ queries introduced by {\bf Exploit-SI} (contains $d$) as follows. The signs that need to be flipped are colored in red.
\begin{eqnarray*}
\begin{array}{c}
\theta=4\\
\begin{array}{|c|c|c|c|c|}\hline
B& S(\Delta_{d}) &\mbox{ Server } $n$ \\ \hline
4 & 1(4)& a_{j_5}-b_{j_2}+c_{j_1}-{\color{brown}d_{*}}  \\  \cline{2-4}
 &2(3)&{\color{red}-}a_{j_6}{\color{red}+}b_{j_3}+{\color{brown}d_{*}}-e_{j_1}  \\
 &2(3)&{\color{red}-}a_{j_7}{\color{red}+}b_{j_4}+{\color{brown}d_{*}}-f_{j_1} \\
  &2(3)&{\color{red}-}a_{j_8}{\color{red}+}c_{j_3}+{\color{brown}d_{*}}-e_{j_2} \\
 &2(3)&{\color{red}-}a_{j_9}{\color{red}+}c_{j_4}+{\color{brown}d_{*}}-f_{j_2} \\
 &2(3)&{\color{red}-}b_{j_8}{\color{red}+}c_{j_6}+{\color{brown}d_{*}}-e_{j_5} \\
 &2(3)&{\color{red}-}b_{j_9}{\color{red}+}c_{j_7}+{\color{brown}d_{*}}-f_{j_5} \\ \cline{2-4}
  &3(2)&{\color{red}}a_{j_{10}}-{\color{brown}d_{*}}{\color{red}-}e_{j_4}{\color{red}+}f_{j_3} \\
 &3(2)&{\color{red}}b_{j_{10}}-{\color{brown}d_{*}}{\color{red}-}e_{j_7}{\color{red}+}f_{j_6} \\
 &3(2)&{\color{red}}c_{j_{10}}-{\color{brown}d_{*}}{\color{red}-}e_{j_9}{\color{red}+}f_{j_8}\\ \hline
\end{array}
\end{array}
\end{eqnarray*}
}

Note that $\sigma_i$ appears in all message variables with symbol index $i$, so $\sigma_i$ might be flipped multiple times and we need to make sure that $\sigma_i$ is flipped consistently, i.e., the sign flipping rule either changes or does not change the signs of all variables with the same index. This is indeed true, proved as follows. Note that we flip the signs depending on whether the sub-block index is even or odd and if the variables are to the left or right of $u_\theta$. This means, for variables in two consecutive sub-blocks, the variables to the left of $u_\theta$ in one sub-block and the variables to the right of $u_\theta$ in the other sub-block are simultaneously flipped or unflipped. So it suffices to show that
all variables with the same index are
\begin{itemize}
\item either in the same sub-block, and all are on the same side of $u_\theta$,
\item or in two consecutive sub-blocks, but are on different sides of $u_\theta$. 
\end{itemize}

{\it Example 4 (continued): Referring to the table above, consider all variables with index $j_1$, i.e., $c_{j_1}, e_{j_1}, f_{j_1}$. $c_{j_1}$ is in sub-block 1 and is to the left of $d$. $e_{j_1}, f_{j_1}$ are in sub-block 2 and are to the right of $d$. Further, the signs of $c_{j_1}, e_{j_1}, f_{j_1}$ are all unflipped. As another example, consider all variables with index $j_{10}$, i.e., $a_{j_{10}}, b_{j_{10}}, c_{j_{10}}$. They are all in sub-block 3 and their signs are all unflipped. One more example: all variables with index $j_6$, $a_{j_6}, c_{j_6}, f_{j_6}$. $a_{j_6}, c_{j_6}$ are in sub-block 2 and are to the left of $d$. $f_{j_6}$ is in sub-block 3 and is to the right of $d$. The signs of $a_{j_6}, c_{j_6}, f_{j_6}$ all need to be flipped.
}

We now find variables with the same symbol index, say $\#$. From index assignment, we know that all occurrences of symbol index $\#$ are in queries that contain the same $m - 1$ (distinct) variables ($u_\theta$ included). 
Suppose the message indices of these $m-1$ variables are $i_{[1:m-2]} \cup \theta$, and let the remaining $M-(m-1)$ message indices be denoted by $r_{[1:M-(m-1)]}$. Assume that $i_1 < i_2 \cdots < i_j < u_\theta < u_{j+1} \cdots < u_{i_{m-2}}$.
Then the symbol index $\#$ appears in queries   
\begin{eqnarray}
&&{\color{blue}\pm u_{r_1}(\#)} \pm u_{i_1}() \pm \cdots \pm u_{i_j}() \pm {\color{brown}u_{\theta}()} \pm u_{i_{j+1}}() \pm \cdots \pm u_{i_{m - 2}}() \notag\\
&&~~\vdots \notag\\
&&\pm u_{i_1}() \pm \cdots \pm u_{i_j}() \pm {\color{brown}u_{\theta}()} \pm u_{i_{j+1}}() \pm \cdots \pm u_{i_{m - 2}}() \pm {\color{blue}u_{r_{M-(m-1)}}(\#)} \label{q1}
\end{eqnarray}
where $\pm$ represents either `$+$' or `$-$', determined by sign assignment.
These $M - (m - 1)$ variables $u_{r_l}, l\in[1:M-(m-1)]$ can be divided into two sets (one set could be empty), where
\begin{itemize}
\item the first set are those $u_{r_l}$ where $r_l < \theta$
\item and the second set are those $u_{r_l}$ where $r_l > \theta$
\end{itemize}
So the variables in the first set are to the left of $u_\theta$ and the variables in the second set are to the right of $u_\theta$. Further, the two sets are in consecutive sub-blocks because $\Delta_{u_\theta}$ only differs by 1. Therefore the sign flipping rule is consistent and the privacy proof is complete.

{\it Example 4 (continued): Suppose we want to find all variables with index $\# = j_1$. They appear in queries that contain $a, b, d$.
The queries in (\ref{q1}) are
\begin{eqnarray*}
\begin{array}{ccc}
a_{j_5}-b_{j_2}+{\color{blue}c_{j_1}}-{\color{brown}d_{*}}\\
{\color{red}-}a_{j_6}{\color{red}+}b_{j_3}+{\color{brown}d_{*}}-{\color{blue}e_{j_1}}\\
{\color{red}-}a_{j_7}{\color{red}+}b_{j_4}+{\color{brown}d_{*}}-{\color{blue}f_{j_1}}
\end{array}~~
\end{eqnarray*}
The 3 variables with index $\# = j_1$ are $c_{j_1}, e_{j_1}, f_{j_1}$ (colored in blue). The first set contains $c_{j_1} (< d)$ (in sub-block 1) and the second set contains $e_{j_1}, f_{j_1} (> d)$ (in sub-block 2). 
As another example, suppose we want to find all variables with index $\# = j_{10}$. The queries in (\ref{q1}) are
\begin{eqnarray*}
\begin{array}{ccc}
{\color{red}}{\color{blue}a_{j_{10}}}-{\color{brown}d_{*}}{\color{red}-}e_{j_4}{\color{red}+}f_{j_3}\\ 
{\color{red}}{\color{blue}b_{j_{10}}}-{\color{brown}d_{*}}{\color{red}-}e_{j_7}{\color{red}+}f_{j_6}\\
{\color{red}}{\color{blue}c_{j_{10}}}-{\color{brown}d_{*}}{\color{red}-}e_{j_9}{\color{red}+}f_{j_8}\\
\end{array}~~
\end{eqnarray*}
The $3$ variables with index $\# = j_{10}$ are $a_{j_{10}}, b_{j_{10}}, c_{j_{10}} (< d)$. They all belong to the first set (sub-block 3).
One more example: find all variables with index $\# = j_6$. 
The queries in (\ref{q1}) are
\begin{eqnarray*}
\begin{array}{ccc}
{\color{red}-}{\color{blue}a_{j_6}}{\color{red}+}b_{j_3}+{\color{brown}d_{*}}-e_{j_1}\\
{\color{red}-}b_{j_8}{\color{red}+}{\color{blue}c_{j_6}}+{\color{brown}d_{*}}-e_{j_5}\\
{\color{red}}b_{j_{10}}-{\color{brown}d_{*}}{\color{red}-}e_{j_7}{\color{red}+}{\color{blue}f_{j_6}}
\end{array}~~
\end{eqnarray*}
The $3$ variables with index $\# = j_6$ are $a_{j_6}, c_{j_6}, f_{j_6}$. The first set contains $a_{j_6}, c_{j_6} (< d)$ (in sub-block 2) and the second set contains $f_{j_6} (> d)$ (in sub-block 3). 
}

\section{Conclusion}
Motivated by privacy concerns in distributed computing, we introduce the private computation problem where a user wishes to compute a desired function of datasets stored at distributed servers without disclosing any information about the function that he wishes to compute to any individual server. The private computation problem may be seen as a generalization of the PIR problem by allowing dependencies among messages. We characterize in Theorem \ref{thm} the capacity of private computation for arbitrary $N$ servers, arbitrary $K$ independent datasets, and arbitrary $M$ linear combinations of the $K$ independent datasets as the possible functions. Surprisingly, this capacity turns out to be identical to the capacity of PIR with $N$ servers and $K$ independent messages. Thus, there is no loss in capacity from the expansion of possible messages to include arbitrary linear combinations. 


Going beyond linear-combinations, we show in Theorem \ref{thm2} that in the asymptotic limit where the number of independent datasets $K\rightarrow\infty$, the capacity of private computation is not affected by allowing non-linear functions into the set of functions that may be computed by the user, provided the symbol-wise entropy of each of these functions is no more than the entropy of a symbol from a dataset. 

In the non-asymptotic regime, the capacity of private computation with arbitrary (non-linear) functions is an interesting direction for future work. Along these lines, let us conclude with the following two observations. The first observation is a general achievability argument for private computation. Consider the most general setting, where we allow the $M$ messages to be arbitrarily dependent and even the entropies of the message symbols are allowed to be different for different messages. Suppose  each message $W_m, m\in[1:M]$ is made of $L$ symbols $W_m=(W_{m,1}, W_{m,2}, \cdots, W_{m,L})$. While the messages may have arbitrary dependencies, the sequence of symbols is generated i.i.d. in $l$, i.e., for all $l\in[1:L]$, the symbols $(W_{1,l}, W_{2,l}, \cdots, W_{M,l})\sim(w_1, w_2, \cdots, w_M)$.  We have \begin{eqnarray}
H(W_1, \cdots, W_M) &=& LH(w_1, \cdots, w_M) \label{wl2}\\
H(W_m) &=& LH(w_m), m \in [1:M] \label{wl1}
\end{eqnarray}
Symbols from different messages may not have the same entropy, i.e., we allow the possibility that $H(w_i)\neq H(w_j)$.
In this  general setting, the private computation rate of $R = \frac{H_{\min}}{H_{\max}} (1 - \frac{1}{N})$ is always achievable, (although not optimal in general) where $H_{\max} = \max(H(w_1), H(w_2), \cdots, H(w_M))$ and $H_{\min} = \min(H(w_1), H(w_2), \cdots, H(w_M))$. Just like the achievability argument for Theorem \ref{thm2}, the general achievability claim follows essentially from \cite{Sun_Jafar_SPIR}. For example, suppose $N=2$. First we compress each message separately into $H_{\max}$ bits per message symbol. This is possible because $\forall m\in[1:M], H(w_m)\leq H_{\max}$. Then, in order to retrieve the $i^{th}$ bit of the compressed desired message, $W_{\theta,i}$, the user  requests from Server $1$, the linear combination $\sum_{m=1}^{M}c_mW_{m,i}$ and from Server $2$, the linear combination $\sum_{m=1}^{M}c_mW_{m,i}+W_{\theta,i}$, where $c_m$ are i.i.d. uniform binary coefficients generated privately by the user and all operations are over $\mathbb{F}_2$. Adding the  answers received from the two servers, allows the user to recover $W_{\theta,i}$. The total number of bits downloaded is $2H_{\max}$, while the number of desired bits retrieved is at least $H_{\min}$. Thus, the rate achieved is at least $\frac{H_{\min}}{2H_{\max}}=\frac{H_{\min}}{H_{\max}} (1 - \frac{1}{N})$ for $N=2$. Similarly, following the approach of \cite{Sun_Jafar_SPIR}, the rate $\frac{H_{\min}}{H_{\max}} (1 - \frac{1}{N})$ is achieved for arbitrary $N$.


The second observation is the capacity characterization for an elemental case where we have $M=2$ arbitrarily correlated messages and $N$ servers. Again consider the  general setting with arbitrary dependencies and without loss of generality,  suppose $H(w_1) \geq H(w_2)$. In this case, the capacity is $C = \frac{NH(w_2)}{H(w_1, w_2) + (N-1)H(w_1)}$.

The converse is proved as follows. From Fano's inequality, we have
\begin{eqnarray}
&& LH(w_1) \notag\\
&\overset{(\ref{wl1})}{=}& H(W_1)\\
&\overset{(\ref{corr})}{=}& I(W_1; A_1^{[1]}, Q_1^{[1]}, \cdots, A_N^{[1]}, Q_N^{[1]}) + o(L) \\
&\overset{(\ref{qmind})}{=}& I(W_1; A_1^{[1]}, \cdots, A_N^{[1]} | Q_1^{[1]}, \cdots, Q_N^{[1]}) + o(L) \\
&=& H(A_1^{[1]}, \cdots, A_N^{[1]} | Q_1^{[1]}, \cdots, Q_N^{[1]})  - H(A_1^{[1]}, \cdots, A_N^{[1]} | W_1, Q_1^{[1]}, \cdots, Q_N^{[1]}) + o(L) \\
&\overset{(\ref{eta_def})}{\leq}& D - H(A_1^{[1]} | W_1, Q_1^{[1]}, \cdots, Q_N^{[1]}) +o(L)\\
&\overset{}{=}& D - H(A_1^{[1]} | W_1, Q_1^{[1]}) + o(L)\label{dd1} \\
&\overset{(\ref{privacy})}{=}& D - H(A_1^{[2]} | W_1, Q_1^{[2]}) +o(L) \label{dd4}
\end{eqnarray}
where (\ref{dd1}) follows from that $H(A_1^{[1]} | W_1, Q_1^{[1]}, \cdots, Q_N^{[1]}) = H(A_1^{[1]} | W_1, Q_1^{[1]})$, proved as follows.
\begin{eqnarray}
&& I(A_1^{[1]}; Q_2^{[1]}, \cdots, Q_N^{[1]} | W_1, Q_1^{[1]}) \notag\\
&\leq& I(A_1^{[1]}, W_2; Q_2^{[1]}, \cdots, Q_N^{[1]} | W_1, Q_1^{[1]}) \\
&=& I(W_2; Q_2^{[1]}, \cdots, Q_N^{[1]} | W_1, Q_1^{[1]}) + I(A_1^{[1]}; Q_2^{[1]}, \cdots, Q_N^{[1]} | W_1, W_2, Q_1^{[1]}) \\
&\overset{(\ref{ans_det})}{=}& I(W_2; Q_2^{[1]}, \cdots, Q_N^{[1]} | W_1, Q_1^{[1]}) \\
&\overset{}{\leq}& I(W_2, W_1; Q_2^{[1]}, \cdots, Q_N^{[1]} | Q_1^{[1]}) \\
&\overset{}{\leq}& I(W_2, W_1; Q_1^{[1]}, \cdots, Q_N^{[1]}) \\
&\overset{(\ref{qmind})}{=}& 0 \label{aa1}
\end{eqnarray}
By a similar argument, we have 
\begin{eqnarray}
I(A_1^{[2]}; Q_2^{[2]}, \cdots, Q_N^{[2]} | W_1, Q_1^{[2]}) &=& 0 \label{aa2} \\
I(A_n^{[2]}; Q_1^{[2]}, \cdots, Q_{n-1}^{[2]}, Q_{n+1}^{[2]}, \cdots, Q_N^{[2]} | W_1, Q_n^{[2]}) &=& 0 \label{aa3}
\end{eqnarray}
Next, from (\ref{dd4}), by symmetry, we have
\begin{eqnarray}
LH(w_1) &\leq& D - H(A_n^{[2]} | W_1, Q_n^{[2]}) + o(L), \forall n \in [2:N] \label{dd2}
\end{eqnarray}
Adding (\ref{dd4}) and (\ref{dd2}) for all $n \in [2:N]$, we have
\begin{eqnarray}
NLH(w_1) &\leq& ND - \sum_{n=1}^N H(A_n^{[2]} | W_1, Q_n^{[2]})  + o(L)\\
&\overset{(\ref{aa2})(\ref{aa3})}{=}& ND - \sum_{n=1}^N H(A_n^{[2]} | W_1, Q_1^{[2]}, \cdots, Q_N^{[2]}) + o(L) \\
&\leq& ND - H(A_1^{[2]}, \cdots, A_N^{[2]} | W_1, Q_1^{[2]}, \cdots, Q_N^{[2]}) + o(L) \\
&\overset{(\ref{corr})}{=}& ND - H(A_1^{[2]}, \cdots, A_N^{[2]}, W_2 | W_1, Q_1^{[2]}, \cdots, Q_N^{[2]}) + o(L) \\
&\overset{}{\leq}& ND - H(W_2 | W_1, Q_1^{[2]}, \cdots, Q_N^{[2]}) + o(L) \\
&\overset{(\ref{qmind})}{\leq}& ND - H(W_2 | W_1) + o(L) \\
&\overset{(\ref{wl2})(\ref{wl1})}{=}& ND - L H(w_2 | w_1) + o(L) \label{ddd} \\
\Longrightarrow ~~ R &=& \frac{H(W_2)}{D} \leq \lim_{L \rightarrow \infty} \frac{LH(w_2)}{\frac{1}{N}\left(NLH(w_1) + LH(w_2|w_1) + o(L)\right)} \\
&=& \frac{NH(w_2)}{H(w_1, w_2) + (N-1)H(w_1)}
\end{eqnarray}
The converse proof is thus complete.

The achievability is based on $\PIRB$. Consider $L=N^2$ symbols of each message at a time. The user privately generates a random permutation over $[1:N^2]$, and applies the same permutation to both messages, taken $N^2$ symbols at a time. Denote this random permutation of the $N^2$ symbols from $W_1$ as $a_1, a_2, \cdots, a_{N^2}$. Similarly, the corresponding random permutation of the $N^2$ symbols from $W_2$ is denoted as $b_1, b_2. \cdots, b_{N^2}$. Note that only symbols with the same index are correlated. Without loss of generality, suppose $W_2$ is desired, and consider the queries generated according to $\PIRB$. 
\begin{eqnarray*}
\begin{array}{c}
\theta = 2\\
\begin{array}{|c|c|c|c|c|}\hline
\mbox{\small Server 1} &\mbox{\small Server 2} &\cdots &\mbox{\small Server $N$} \\ \hline
a_1, b_1&a_2,b_2&\cdots&a_{N}, b_{N}\\
a_2+b_{N+1}& a_1+b_{2N} & \cdots & a_1+b_{N^2-N+2}\\
\vdots&\vdots&\ddots&\vdots\\
a_N+b_{2N-1} & a_N+b_{3N-2} &\cdots&a_{N-1} + b_{N^2}\\
\hline
\end{array} 
\end{array}
\end{eqnarray*}
In order to send $(a_1,b_1)$, Server $1$ needs only $H(w_1,w_2)$ bits. Note that optimal compression requires long sequences, so the scheme operates over $LN^2$ symbols each of $W_1$ and $W_2$, for large $L$, so that $a_1$ is a sequence of $L$ symbols from $W_1$, and $b_1$ is the corresponding sequence of $L$ symbols from $W_2$, and optimal compression is possible as $L\rightarrow\infty$. Thus, for $(a_1, b_1)$ the server sends $LH(w_1,w_2)+o(L)$ bits. For $a_2+b_{N+1}$, the key is that the server first  compresses the $L$ symbols of $a_2$, and the $L$ symbols of $b_{N+1}$, separately, each into $LH(w_1)+o(L)$ bits. This is possible because $H(w_1)\geq H(w_2)$. And then the server sends the sum of the compressed bits, for a total of $LH(w_1)+o(L)$ bits. Each  $2$-sum $a+b$ is compressed similarly. Thus, the total download from Server $1$ is $LH(w_1,w_2)+L(N-1)H(w_1)+o(L)$ bits. The total download from all servers is $N$ times that number of bits. The total number of desired bits retrieved is $LN^2H(w_2)$. Therefore, the rate achieved is  $\lim_{L\rightarrow\infty}LN^2H(w_2)/N(LH(w_1,w_2)+L(N-1)H(w_1)+o(L))=NH(w_2)/(H(w_1,w_2)+(N-1)H(w_1))$, and the capacity for this case is settled. Finding the capacity for $3$ or more dependent messages with arbitrary dependencies is the next immediate open problem for future work.


\bibliographystyle{IEEEtran}
\bibliography{Thesis}
\end{document}